\documentclass[10pt,twocolumns,apj]{emulateapj}

\usepackage{afterpage}
\newcommand{\D}{\displaystyle}
\newcommand\eex[1]{\mbox{$\times 10^{#1}$}}     % #1 x 10^#2
\newcommand\eez[1]{\mbox{$10^{#1}$}}            % just 10^#2
\newcommand\hi{\ion{H}{1}}

\newcommand\halpha{\mbox{H$\alpha$}}
\newcommand\vdev{\mbox{$v_{\rm dev}$}}
\newcommand\vhel{\mbox{$v_{\rm helio}$}}
\newcommand\lsun{\hbox{L$_{\odot}$}}
\newcommand\msun{\hbox{$M_{\odot}$}}
\newcommand\mdot{\hbox{$\dot M$}}
\newcommand\zsun{\mbox{$Z_{\odot}$}}
\newcommand\V{\mbox{$V$}}%% V filter
\newcommand\rsloan{\hbox{$r^\prime$}}
\newcommand\kps{\mbox{${\rm km~s^{-1}}$}}
\newcommand\cm{\mbox{${\rm cm^{-2}}$}}
\newcommand\nh{\mbox{$N_{\rm HI}$}}
\newcommand\mhi{\mbox{$M_{\rm HI}$}}
\newcommand\mtot{\mbox{$M_{\rm tot}$}}
\newcommand{\pv}{\mbox{$p$-$v$}}
\newcommand{\bc}{\begin{center}}
\newcommand{\ec}{\end{center}}

\slugcomment{Accepted by the Astrophysical Journal}

\shorttitle{High-Velocity Clouds in M 83}
\shortauthors{Miller et al.}

\begin{document}

\title{High-Velocity Clouds in the Nearby Spiral Galaxy M 83}

\author{Eric D. Miller\altaffilmark{1},
Joel N. Bregman\altaffilmark{2},
and Bart P. Wakker\altaffilmark{3}}

\altaffiltext{1}{Kavli Institute for Astrophysics and Space Research,
Massachusetts Institute of Technology, Cambridge, MA 02139;
milleric@mit.edu}
\altaffiltext{2}{Department of Astronomy, University of Michigan, 
Ann Arbor, MI 48105}
\altaffiltext{3}{Department of Astronomy, University of Wisconsin, 
Madison, WI 53706}

\begin{abstract}

We present deep \hi\ 21-cm and optical observations of the face-on spiral
galaxy M 83 obtained as part of a project to search for high-velocity
clouds (HVCs) in nearby galaxies.  Anomalous-velocity neutral gas is
detected toward M 83, with 5.6\eex{7} \msun\ of \hi\ contained in a disk
rotating 40--50 \kps\ more slowly in projection than the bulk of the gas.
We interpret this as a vertically extended thick disk of neutral material,
containing 5.5\% of the total \hi\ within the central 8 kpc.  Using an
automated source detection algorithm to search for small-scale \hi\
emission features, we find eight distinct, anomalous-velocity \hi\ clouds
with masses ranging from 7\eex{5} to 1.5\eex{7} \msun\ and velocities
differing by up to 200 \kps\ compared to the \hi\ disk.  Large on-disk
structures are coincident with the optical spiral arms, while unresolved
off-disk clouds contain no diffuse optical emission down to a limit of 27
\rsloan\ mag per square arcsec.  The diversity of the thick \hi\ disk and
larger clouds suggests the influence of multiple formation mechanisms, with
a galactic fountain responsible for the slowly-rotating disk and on-disk
discrete clouds, and tidal effects responsible for off-disk cloud
production.  The mass and kinetic energy of the \hi\ clouds are consistent
with the mass exchange rate predicted by the galactic fountain model.  If
the HVC population in M 83 is similar to that in our own Galaxy, then the
Galactic HVCs must be distributed within a radius of less than 25 kpc.

\end{abstract}

\keywords{galaxies: individual (M 83) --- galaxies: ISM --- galaxies: kinematics and dynamics --- radio lines: galaxies}

%%%%%%%%%%%%%%%%%%%%%%%%%%%%%%%%%%%%%%%%%%%%%%%%%%%%%%%%%%%%%%%%%%%%%%%%
% SECTION -- Introduction
%%%%%%%%%%%%%%%%%%%%%%%%%%%%%%%%%%%%%%%%%%%%%%%%%%%%%%%%%%%%%%%%%%%%%%%%
\section{INTRODUCTION}

Our understanding of the gaseous content of our Galaxy has advanced greatly
in recent years as instruments sensitive to a variety of wavebands have
become available.  One component of the Galaxy's gaseous neighborhood
remains poorly understood: the ensemble of high-velocity clouds (HVCs).
Initially discovered in \citeyear{Mulleretal63} by
\citeauthor*{Mulleretal63}, HVCs are clouds of \hi\ gas that deviate in
velocity from the differentially rotating disk of gas by 70--300 \kps.
These structures have been studied extensively in \hi\ emission, \halpha\
emission, and absorption against background point sources, with several
published catalogs detailing their observed characteristics
\citep{WvW91,WvW97,Wakker01,Putmanetal2002}.

A cursory inspection of the HVC catalogs reveals a tremendous variation in
size, morphology, and velocity; this variation has led to a number of
formation scenarios that predict different physical properties.  One
favored model is the Galactic fountain, a cyclic phenomenon whereby hot ($T
\sim \eez{6}$ K) gas is ejected into the lower halo ($z \lesssim 10$ kpc)
by supernova explosions \citep{ShapiroField76,Bregman80}.  Here the gas
cools and loses its buoyancy, free-falling back to the disk in sheets at
mainly negative velocities approaching 100 \kps.  A second scenario
suggests accretion of close companion galaxies, either through tidal or
ram-pressure stripping as a companion moves through the hot, extended
Galactic halo.  This is undoubtedly the origin of the Magellanic Stream
\citep{Mathewson74}, and there is recent evidence for gaseous remnants of
satellite galaxies \citep{Lockman2003,Putmanetal2004}.  Other scenarios
propose alternate extragalactic origins, such as accreted intergalactic
medium (IGM) material \citep{Oort66,Oort70,Oort81} or discrete remnants of
galaxy formation \citep{Oort66,Blitz99,BraunBurton99}.  This last model
attributes the most compact HVCs (CHVCs) to an ensemble of primordial,
dark-matter-dominated ``minihalos'' streaming into the Local Group along
filaments of moderate overdensity.  Such clouds would have a typical
distance of $\sim$ 750 kpc, and total mass of $\sim \eez{8}$--$\eez{9}$
\msun\ per cloud (assuming $\mhi = 0.1 \mtot$).  Modeling by
\citet{SternbergMcKeeWolfire02} and simulations of CDM halo evolution by
\citet{Klypinetal99} and \citet{Mooreetal99} support a more local
distribution of $\sim$ 150 kpc.  The clouds need not be dominated by dark
matter, as uneven CDM halo cooling can result in fragmentation and produce
a population of moderately-ionized, pressure-confined clouds embedded in a
hot corona \citep{MallerBullock04}.  

\begin{deluxetable*}{lrrrrr}
\tabletypesize{\normalsize}
\tablewidth{0pt}
\tablecaption{Predictions of HVC Production Scenarios
     \label{tab:scenarios}}
\tablehead{
\colhead{scenario} & 
\colhead{distance} & 
\colhead{$|\vdev|$} & 
\colhead{size} & 
\colhead{$M_{\rm cloud}$} &
\colhead{$Z$} \\
\colhead{} & 
\colhead{(kpc)} & 
\colhead{(\kps)} & 
\colhead{(kpc)} & 
\colhead{(\msun)} & 
\colhead{(\zsun)}
}
\startdata
IGM infall          & $< 3$        & $< 150$  & $< 1$     & $<$\eez{4} 
 & 0.1--0.3 \\
Galactic fountain   & $< 10$       & $< 150$  & $< 1$  & \eez{4}--\eez{5} 
 & $\ge$ 1 \\
companion stripping & 5--100       & $< 300$  & $\sim 1$  & \eez{5}--\eez{6} 
 & 0.1--1 \\
circumgalactic warm clouds & $\sim 150 $ & $< 300$ & $\sim 1$ & \eez{6}-\eez{7} 
 & 0.1--0.3 \\
circumgalactic DM halos & $\sim 150 $  & $< 300$  & $\sim 1$ & \eez{6}-\eez{7} 
 & 0.1--0.3 \\
Local Group DM halos    & $\sim 750 $  & $< 300$  & 1--10 & \eez{8} 
 & 0.1--0.3 \\
\enddata
\tablecomments{\footnotesize{Sizes and masses shown are typical values based on the
median observed parameters of the HVC ensemble.  Masses for the first three
scenarios assume \hi\ is the primary constituent of the cloud, while the
others assume an neutral gas fraction of 0.1 (warm clouds) or \hi\ to dark
matter ratio of 0.1 (dark matter halos).  Velocities for the Galactic
fountain and IGM infall models assume an adiabatic corona of fully ionized
gas with $T \sim \eez{6}$ K.}} 
\end{deluxetable*}

One key discriminator between these production mechanisms is distance, yet
this parameter is difficult to determine since HVCs contain no visible
stars.  The most direct method of searching for HVC absorption against
background halo stars has met with limited success until very recently,
yielding upper
limits on the distances to five HVCs and eight intermediate-velocity clouds
\citep[IVCs;][]{Wakker01}.  The upper bounds to the vertical $z$ heights of
these HVCs range from 0.2 kpc to 7 kpc, and the corresponding mass limits
range from about 1 \msun\ for a very small HVC to 2\eex{6} \msun\ for
Complex A.  
Recent work has placed stronger constraints on an additional four HVCs,
placing them at distances of $\sim$ 5--15 kpc
\citep{Thometal2006,Wakkeretal2007,Wakkeretal2008}.
In particular, \citet{Wakkeretal2008} infer a distance of 3.7--11.2 kpc for
Complex C, suggesting an \hi\ mass of 3--14\eex{6} \msun.

Indirect methods have also been used to determine HVC distances,
such as constraining the degree of ionization on the cloud surface.  It is
difficult to model the strength of the UV radiation field leaking out of
the Milky Way, although there are indications that the field is sufficient
to ionize at least the surfaces of an HVC \citep{BregmanHarrington86,
Bland-HawthornMaloney97}.  Several groups have identified ionized ``skins''
in a number of HVCs
\citep{WeinerWilliams96,Tufteetal98,Bland-Hawthornetal98}, and observations
by \citet{Putmanetal2003} of \halpha\ emission in 25 HVCs limit their
distances to $5 < z < 40$ kpc.  

While recent progress has been made, large gaps remain in our
understanding of Galactic HVCs, the nature of the gaseous Galactic halo,
and the relationship between such a halo and the thin, star-forming disk.
The various HVC production scenarios predict different consequences for
galaxy formation and evolution.  In Table \ref{tab:scenarios}, we outline
the predictions of the scenarios described above in terms of observables
(size, velocity) and physical properties (mass, metallicity).  Our vantage
point remains an impediment to discriminating between these scenarios, as
we are embedded within the very medium we attempt to disentangle.  The
superposition of material in position and velocity complicates the
analysis, and chance locations of statistically anomalous sources (such as
the Magellanic Stream and large HVC complexes) bias the conclusions.

These problems are largely overcome by observing a large sample of external
galaxies, constraining the fundamental properties of their diffuse halos,
and extrapolating to the Milky Way.  Until recently, extragalactic \hi\
studies have concentrated on the structure of the kinematically cold disk,
and the discovery of anomalous-velocity material has been serendipitous.
Early evidence of anomalous \hi\ appeared in the face-on galaxies M 101
\citep{vanderHulstSancisi88,Kamphuisetal91}, NGC 628
\citep{KamphuisBriggs92}, and NGC 6946 \citep{KamphuisSancisi93}, with
detections of a few clouds of mass $\mhi > \eez{7}$ \msun.  Integrated \hi\
profiles from a sample of 14 face-on galaxies show evidence for
high-velocity gas in 10 of the systems \citep{Schulmanetal1994}.  More
recent observations of inclined galaxies have uncovered anomalous \hi\ gas
in every target imaged with sufficient depth.  Deep radio synthesis
observations of NGC 891
\citep{SwatersSancisivanderHulst97,Oosterlooetal2007}, NGC 2403
\citep{Fraternalietal02a}, NGC 4559 \citep{Barbierietal2005}, NGC 253
\citep{Boomsmaetal2005}, 
and NGC 6946 \citep{Boomsmaetal2008}
have revealed
vertically extended, slowly rotating ``thick'' disks of \hi, suggestive of
Galactic fountain activity.  Modeling of a subset of these data confirm
this conclusion, although some amount of IGM infall is required to prevent
escape of the fountain gas and to produce the gas kinematics that are
observed \citep{FraternaliBinney2008}.  A GBT study of M 31 has uncovered
for the first time a population of discrete HVC-like \hi\ clouds in an
external galaxy \citep{Thilkeretal04}.  These objects are all within 50 kpc
of the galaxy.  Surveys of galaxy groups have so far failed to detect HVC
analogs without optical counterparts \citep[e.g.,][and references
therein]{Zwaan01,Pisanoetal2007}.  Extrapolation by \citet{Pisanoetal2007}
of results from six Local Group analogs suggest the Galactic HVCs all lie
within 90 kpc of the Galaxy.

In this paper, we describe the first results of a search for HVCs in nearby
external spiral galaxies.  We concentrate on face-on galaxies, since the
line-of-sight velocity of the rotating \hi\ disk is minimized and the
parameter space available for HVC searching is maximized.  These targets
provide a necessary complement to the numerous inclined spirals which have
been recently observed, as detailed above.  By combining the \hi\ data with
deep optical surface photometry, we can constrain the amount of starlight
in any anomalous \hi\ clouds we find.

The next section summarizes the observations and reduction of both the \hi\
and broad-band optical data. In Section 3, we discuss features of the bulk
\hi\ and optical disks.  In Section 4, the cold \hi\ disk is modeled and 
subtracted, and extended anomalous \hi\ emission is analyzed.  Section 5
discusses the detection of discrete anomalous \hi\ emission, introducing a
new detection algorithm (further detailed in the Appendix \ref{app:snrch})
and presenting simulations that constrain the effectiveness of this
technique.  The discussion in Section 6 places our results in context of
previous work and predictions.  Our conclusions are summarized in Section
7.

%%%%%%%%%%%%%%%%%%%%%%%%%%%%%%%%%%%%%%%%%%%%%%%%%%%%%%%%%%%%%%%%%%%%%%%%
% SECTION -- Data
%%%%%%%%%%%%%%%%%%%%%%%%%%%%%%%%%%%%%%%%%%%%%%%%%%%%%%%%%%%%%%%%%%%%%%%%
\section{OBSERVATIONS \& DATA REDUCTION}

\begin{deluxetable*}{lrclr}
\tabletypesize{\normalsize}
\tablewidth{0pt}
\tablecaption{M 83 Properties
     \label{tab:m83params}}
\tablehead{\vspace{0pt}}
\startdata
NGC number                    & 5236                 & ~~~ &  Type\tablenotemark{c}                & SBc(s)II \\     
RA (J2000)                  & 13h37m00.8s            &     &  Inclination\tablenotemark{d}         & 24\arcdeg \\
Dec (J2000)        & $-$29\arcdeg51\arcmin59\arcsec  &     &  Pos.\ angle\tablenotemark{d}         & 225\arcdeg \\
$l$                         & 314.6\arcdeg           &     &  $D_{25}$\tablenotemark{b}            & 12\farcm9 \\
$b$                         & 32.0\arcdeg            &     &                              & 17 kpc \\
Distance\tablenotemark{a}            & 4.5 Mpc       &     &  $R_{\rm Holm}$\tablenotemark{e}    & 7\farcm 9 \\
$v_{\rm hel}$\tablenotemark{b}       & 503--516 \kps &     &                              & 10 kpc \\
$B$\tablenotemark{b}                 & 8.2           &     &  Total mass\tablenotemark{e}          & 1--5$\eex{11}$ \msun \\
$M_B$                       & $-$20.1                &     &  HI mass\tablenotemark{e}           & 6.1$\eex{9}$ \msun \\
$A_V$                       & 0.22                   &     &  1\arcmin                    & 1.3 kpc \\
\enddata
\tablenotetext{a}{\footnotesize{\cite{Karachentsevetal02}}}
\tablenotetext{b}{\footnotesize{\cite{RC3}}}
\tablenotetext{c}{\footnotesize{\cite{RSA}}}
\tablenotetext{d}{\footnotesize{\cite{TalbotJensenDufour79}}}
\tablenotetext{e}{\footnotesize{\cite{HuchtBohn81}}}
\end{deluxetable*}

M 83 is a nearby \citep[4.5 Mpc; ][]{Karachentsevetal02}, face-on grand-design
spiral galaxy located in the Centaurus A group.  While the optical,
star-forming disk is small and well defined
\citep{TalbotJensenDufour79}, the \hi\ disk is very
extended, reaching to 6.5 times the optical Holmberg radius with 80\% of
the total \hi\ mass found outside this radius \citep{HuchtBohn81}.  The
\hi\ disk is resolved into distinct rings and arms in this region,
exhibiting a high degree of warping \citep{TilanusAllenM83}.  The basic
properties of M 83 are presented in Table \ref{tab:m83params}.

\subsection{\hi\ 21-cm Data \label{sect:vlaobs}}

The radio data for M 83 were obtained in three separate observations with
the VLA\footnote{The VLA is operated by the National Radio Astronomy
Observatory, a facility of the National Science Foundation operated under
cooperative agreement by Associated Universities, Inc.} 
on 24 and 28 February and 1 March 1999.  To match the beam size with the
scale of the expected emitting regions (1 kpc = 46\arcsec\ at 4.5 Mpc), we
chose to observe in the DnC configuration, in which the northern arm of the
array is in the more extended C configuration.  For sources at southern
declinations such as M 83 (meridian altitude = 26\fdg1 at the VLA's
latitude of $+$34\fdg1), this configuration produces a more symmetric
$uv$-plane coverage pattern and a synthesized beam of about 35\arcsec\
FWHM (0.76 kpc at a distance of 4.5 Mpc).  Each observing run consisted of a
5 min observation of the primary flux calibrator 3C 286, with a scan
integration time of 20 s.  This was followed with a series of $\sim$ 50 min
exposures of M 83 with integration times of 60 s, intermixed with 5 min
exposures of the secondary calibrator 3C 283.  The total time spent on
source was 12.2 h.

The VLA spectral line correlator was configured to observe in the 1420 MHz
frequency band, or ``L-band''.  The correlator configuration and resulting
data parameters are summarized in Table \ref{tab:m83hiparams}.  To allow
detection of \hi\ emission at highly anomalous velocities, we used the largest
bandwidth offered by the correlator, 3.125 MHz.  This corresponds to a
velocity width of 660 \kps, of which only the inner 540 \kps\ were retained
because of edge noise.  With the bandwidth centered on the systemic
velocity of M 83, this enabled detection of features deviating by $\pm 270$
\kps\ from systemic.  We expected the features to have linewidths of
$\sim$~\nolinebreak20--30 \kps\ FWHM, therefore a velocity resolution of
5--10 \kps\ would have been sufficient for this project.  To make the
dataset generally usable to the community, we chose to observe at the full
spectral resolution, with no on-line Hanning smoothing applied, and using
the single IF correlator mode (sensitive only to right-handed circular
polarization).  This resulted in 256 channels with a channel separation of
12.2 kHz (2.58 \kps) and a channel width of 14.6 kHz (3.10 \kps).

The $uv$ calibrator data were edited in a standard iterative way, with the
inner 75\% of the bandpass combined and inspected for obviously bad
samples.  Once all anomalous samples were excised from the calibrator data,
the frequency-independent calibration was applied to the object data, which
were then edited.  The calibration solutions had a maximum closure error of
1\% in amplitude and 1\arcdeg\ in phase.  The (frequency-dependent) bandpass
correction was applied using the secondary calibrator 3C 283 as reference,
again following an iterative editing/calibrating method.  Application of
the bandpass correction produced closure errors of 0.5\% in amplitude and
0.5\arcdeg\ in phase, however because of the limited integration time of
the calibrator, the noise in the final data cube was increased by a factor
of 1.8.  With the editing and calibration completed, the $uv$ data were
Hanning smoothed in velocity space.  Besides improving the signal-to-noise
in individual channels, smoothing with a Hanning kernel reduced the number
of channels from 256 to 128 and made them independent, with channel width
and separation equal at a value of 24.4 kHz (5.17 \kps).  

\begin{deluxetable*}{lrr}
\tabletypesize{\scriptsize}
\tablewidth{0pt}
\tablecaption{VLA Configuration and Data Parameters
     \label{tab:m83hiparams}}
\tablehead{\vspace{0pt}}
\startdata
Dates of observation & \multicolumn{2}{r}{24 Feb, 28 Feb, 1 Mar 1999}  \\
Time on source    & \multicolumn{2}{r}{12.2 hr}  \\
No.~antennas      & \multicolumn{2}{r}{26}  \\
Central freq.     & \multicolumn{2}{r}{1.418 GHz}  \\
Primary beam HPBW & \multicolumn{2}{r}{31\farcm2}  \\
Synth.~beam HPBW  & \multicolumn{2}{r}{$37\farcs2 \times 33\farcs1$}\\
Synth.~beam PA    & \multicolumn{2}{r}{$-72.5\arcdeg$}  \\
1-$\sigma$ noise (per 5.17 \kps\ chan.)          & \multicolumn{2}{r}{0.94 mJy/beam}  \\
1-$\sigma$ HI col. density (per 5.17 \kps\ chan.) & \multicolumn{2}{r}{4.6\eex{18} \cm} \\
1 mJy/beam                     & \multicolumn{2}{r}{0.49 K}          \\
\tableline
\multicolumn{3}{l}{\vspace{-5pt}}                                 \\
                         & Raw data   & Processed/cleaned          \\
\multicolumn{3}{l}{\vspace{-5pt}}                                 \\
\tableline
Bandwidth                & 3.125 MHz  & 2.56 MHz                   \\
                         & 660 \kps   & 540 \kps                   \\
Velocity range           & 183--843 \kps & 230--770 \kps           \\
No.~channels             & 256        & 105                        \\
Channel separation       & 2.58 \kps  & 5.17 \kps                  \\
Channel width            & 3.10 \kps  & 5.17 \kps                  \\
\enddata
\end{deluxetable*}

The ultimate success of any interferometry observation hinges on whether a
reliable map of source intensity in real space can be produced from
incomplete sampling of the coherence function in the $uv$ plane.
At the imaging stage, weighting visibilities
based on their local population density or distance from the tracking
center affects the strength of sidelobe structure.  We applied two
weighting schemes to improve the quality of the principal solution:  first,
the ``robust'' scheme developed by \citet{Briggs95} to produce a mixture of
natural and uniform weighting with a single parameter; and second, a
Gaussian taper function to downweight visibilities in the sparsely-sampled
outer $uv$ plane.  We used the {\tt AIPS IMAGR}\ task \citep{AIPSCookbook}
and applied a variety of values for these two weighting schemes,
empirically determining the best compromise between sidelobe structure,
resolution and sensitivity.  The best compromise among all the
considerations was a ROBUST value of 0 and a Gaussian $uv$-taper with
$r_{.3} = 6$ k$\lambda$, where $r_{.3}$ is the half-width at 30\% of
maximum.  This weighting scheme increased the noise in the final image by a
factor of 1.2 compared to pure natural weighting and no $uv$-taper.  

A dirty image cube was produced from the full $uv$ dataset, and channels
free of \hi\ line emission were identified.  The continuum was subtracted
by fitting a first-order polynomial to 10 line-free channels on each edge
of the band and subtracting this from all channels.  A continuum-free dirty
image cube was produced from this $uv$ dataset, using a cell size of
7\arcsec\ per pixel and only including the central 105 channels to
eliminate edge noise.

To correct for incomplete sampling in the $uv$ plane and reduce the level
of the sidelobes, one typically uses the dirty image to iteratively
construct a model of the true brightness distribution.  This is the
strategy behind the CLEAN algorithm \citep{Hogbom74,Clark80}, under
which the brightness distribution is modeled iteratively as a set of
$\delta$-function ``clean components''.  CLEAN is very
effective at removing sidelobes from point sources, since the clean
components themselves are modeled as points.  Spatially-extended \hi\
emission, however, is poorly modeled by a set of point sources.
Furthermore, while the strength of extended emission is usually relatively
weak compared to a continuum point source, and thus the sidelobes
associated with a single grid point often fall below the noise in a dirty
map, the summation of sidelobes due to complex extended structure can be
significant and difficult to disentangle.

To combat this problem, another method of image deconvolution was developed
by \citet{WakkerSchwarz88}.  Their Multi-Resolution CLEAN (MRC) runs the
CLEAN algorithm, but operates on the point-like and extended sources
separately.  First, the dirty map and beam are smoothed then 
subtracted from the original map and beam to obtain a
full-resolution difference map and beam.  Clean beams are constructed for
the smooth and difference beams, and CLEAN is performed on the
smooth and difference maps separately.  By performing CLEAN on a smoothed
map, two major improvements are made in reproducing extended emission:
the clean beam is closer in size to the extended sources one is
attempting to model; and the signal-to-noise is improved, allowing
detection of fainter sources.  The original resolution is retained (and
point sources are deconvolved) by cleaning the difference map separately
and then combining the two resulting cleaned maps along with the
full-resolution residuals.  A detailed explanation of the method and the
exact scaling parameters used are presented by \citet{WakkerSchwarz88}.

After trying standard CLEAN and MRC and comparing the results, it was
determined that MRC was more effective in removing sidelobes.  Our
procedure was to clean the full region of each channel containing
emission above the 3-$\sigma$ level, then examine the map
for remaining strong sidelobes, which were present in most of
the channels.  The output cube was spatially smoothed to twice the beam
size and binned up by 5 channels, and regions that contained significant
emission were manually delineated with polygons. The cleaning was repeated
with MRC, using the original dirty cube and allowing clean components only
within the specified region of each channel.  Successively deeper cleaning
was performed, down to a threshold of 1-$\sigma$.

\begin{figure}
\plotone{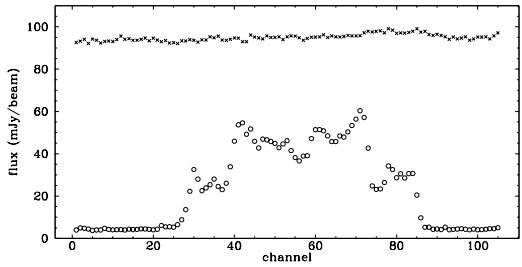}
\caption{The 1-$\sigma$ noise (crosses) and the peak \hi\ flux (circles) in
individual channels of the cleaned M 83 data cube.  The noise levels have
been multiplied by a factor of 100 and placed on the same scale as the peak
flux to facilitate comparison.  The noise is close to constant over the
entire cube, with an RMS of 1.7\% and maximum variations less than 5\% of
the mean.}
\label{fig:m83channoise}
\end{figure}

The 1-$\sigma$ noise level, calculated from the emission-free regions of
each image plane, and other parameters for the cleaned data cube are
summarized in Table \ref{tab:m83hiparams}.  The noise of 0.94 mJy/beam is
larger than the theoretical value of 0.7 mJy/beam for this instrumental
setup and image weighting parameters, and this is likely a result of the
bandpass calibration.  The noise is similar from channel to channel,
varying no more than 5\%, as is shown in Figure \ref{fig:m83channoise}.

\begin{figure*}[p]
\plotone{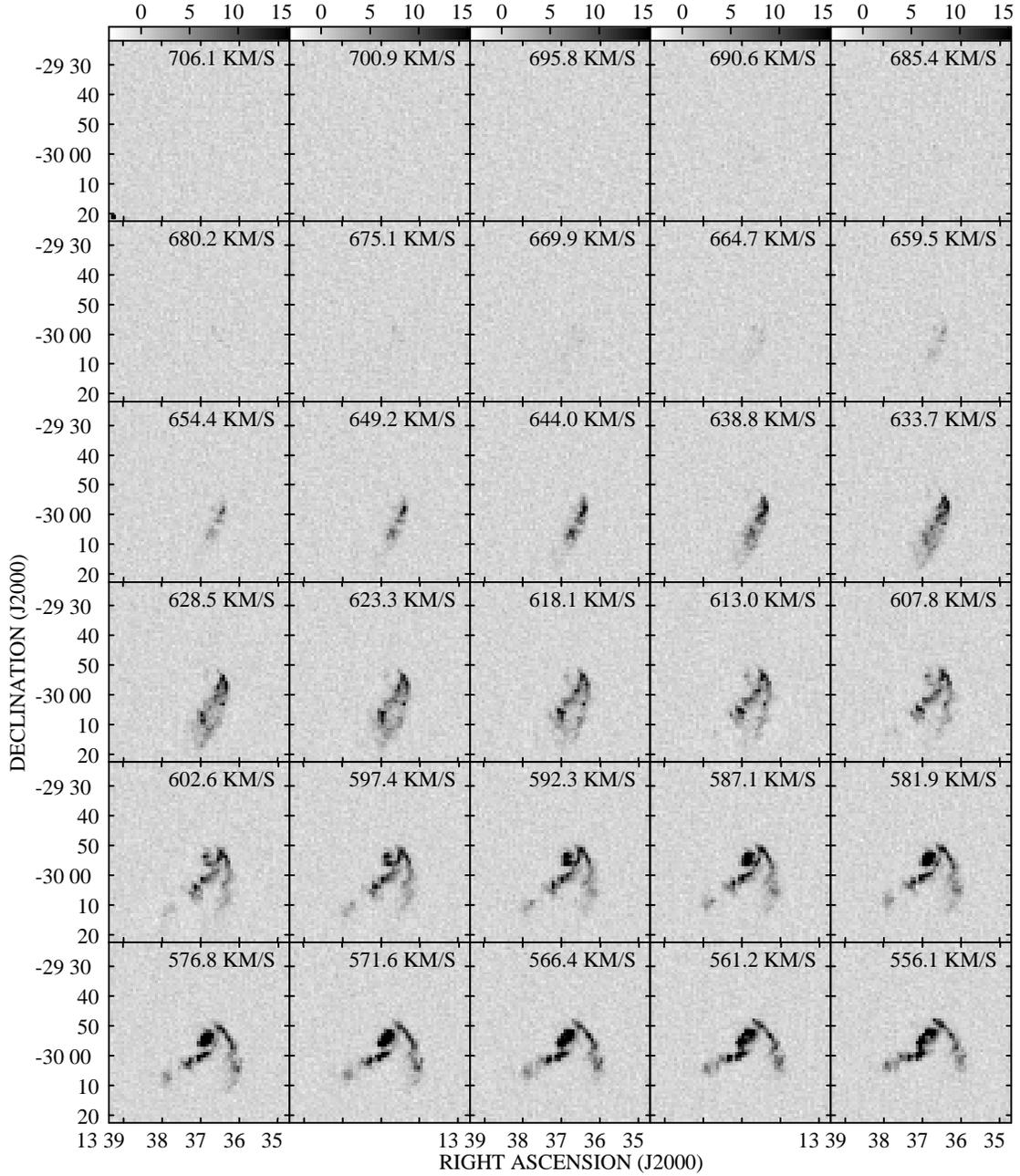}
\caption{M 83 \hi\ channel maps.  Heliocentric velocities are marked in the
upper right of each channel map.  The grayscale ranges from $-3$ to 15 mJy
to show the fainter structure in each channel, therefore most of the bright
features appear saturated.  Only channels with measurable \hi\ emission are
displayed.}
\label{fig:m83chanmaps}
\end{figure*}

To reduce image size and ease further computational analysis, the image
cube was binned $2 \times 2$ in space.  With a final pixel size of
14\arcsec, the $\sim 35$\arcsec\ (FWHM) synthesized beam is still
well-sampled.  Noise was not reduced, however, since neighboring pixels are
correlated.  The channel maps for the cleaned data cube are shown in Figure
\ref{fig:m83chanmaps}.

\subsection{Optical Data}

The optical data for M 83 were taken with the 0.6m Curtis Schmidt telescope
located at Cerro Tololo Inter-American Observatory\footnote{CTIO is part of
the National Optical Astronomy Observatory, which is operated by the
Association of Universities for Research in Astronomy (AURA), Inc., under
cooperative agreement with the National Science Foundation.}
(CTIO) in March 1999.  Use of a thinned SITe 2K CCD chip resulted in a
field of view of nearly 70\arcmin\ with 2.3\arcsec\ square pixels.  To
reduce flat-fielding errors that could be produced by non-repeatable filter
wheel positioning and dislodged dust particles, we used a single Sloan
\rsloan\ filter, keeping the filter wheel unchanged throughout the run.  A
standard observing scheme was followed, with twenty bias frames and six to
eight twilight flats taken each evening, and standard star fields imaged at
a range of airmass throughout the night.  Approximately half of the
telescope time was spent imaging galaxies and half imaging blank sky flats,
with exposure times of 15 minutes each.  The total exposure times were 10
hours on source and 8.75 hours on blank sky.  Dark exposures taken during
one cloudy night indicated no measurable CCD dark current.

The initial reduction procedure followed a standard formula.  The bias and
overscan were removed from all twilight, sky and object frames.  A
first-order flatfield was created for each night from the twilights flats.
This flatfield was applied to the sky flats, which were co-added and
smoothed with a $128\times128$ boxcar to remove stars.  This sky correction
image was normalized and multiplied by the first-order twilight flat field
to produce a master flat field for each night, which was applied to
the M 83 data.  The individual galaxy frames were examined by eye, and
eight of the 40 images contained undesirable features such as strong
scattered light and were rejected.  The remaining frames were
sky-subtracted, aligned, and co-added.  For proper noise estimation in
future analysis, an average sky value of 4423 ADU was added back in to the
co-added frame.  Finally, the poorly-sampled outer regions were clipped,
leaving a final 72\arcmin$\times$69\arcmin\ image with a total integration
time of 8.0 hours.

Photometric calibration was accomplished with observations of standard
stars.  The number of standard stars for the Sloan filter system was
limited at the time of observations, but using a star list provided
by J.A.~Smith (private communication), we identified a number of
standard stars within \citet{Landolt92} fields.  Nightly imaging of these
fields yielded 50 observations of seven standard stars covering a range of
airmasses ($1.07 < \sec{z} < 2.0$) and colors ($0.0 < B-V < 1.0$).  For a
pixel size of 5.31 square arcsec and an effective exposure time of 900 s,
the surface brightness corresponding to 1 ADU pix$^{-1}$ in the final
averaged galaxy frame was $30.11 \pm 0.05$ \rsloan\ magnitudes per square
arcsec.  The average sky value of 4423 ADU corresponds to a surface
brightness of 21.04 \rsloan\ magnitudes per square arcsec, and the
pixel-to-pixel noise in the final image (6.5 ADU) corresponds to 28.1
\rsloan\ magnitudes per square arcsec.  Our 3-$\sigma$ limiting surface
brightness is 26.9 \rsloan\ magnitudes per square arcsec, not accounting
for systematic errors caused by scattered light and large-scale
flatfielding effects.

Foreground stars were removed by modeling the stellar point spread function
(PSF) and masking each star.  Using the DAOPHOT \citep{Stetson87} package
in IRAF\footnote{IRAF is distributed by the National Optical Astronomy
Observatories, which are operated by the Association of Universities for
Research in Astronomy, Inc., under cooperative agreement with the National
Science Foundation.}, we estimated the brightness of all stars in the frame
by searching for 3-$\sigma$ peaks in the image and performing simple
aperture photometry on them.  This procedure produced 38,721 stellar
objects; of these, 11 bright, isolated stars were chosen to construct the
PSF.  The PSF stars had an average (standardized) brightness of \rsloan =
16.3, with one moderately saturated star included to improve the fit in the
wings.  The PSF was constructed from the unsaturated regions of the PSF
stars, using an elliptical Moffat function with $\beta = 2.5$ as the
analytic PSF component plus a single look-up table for the residual PSF
component.  The PSF was scaled to each PSF star and subtracted, leaving a
residual image and revealing previously-unidentified stars in the PSF
stars' wings.  These were fit with the PSF and removed from the original
frame, and a new PSF was constructed.  This procedure was iterated several
times until the companion stars were completely removed and the PSF did not
improve between iterations.

Once the PSF was modeled, it was used to identify and measure all stars in
the field.  Of the stars originally identified by aperture photometry, a
subset of 33,950 were found to have profiles similar to the PSF.  This
amounts to an average of 6.8 field stars per square arcmin, down to a
detection limit of \rsloan = 23.8.  The core of a star becomes saturated at
$\rsloan \sim 14.5$, but by fitting to the unsaturated wings, stars
down to $\rsloan \sim 11$ could be reliably measured.

Stars were removed by masking each one out to the radius at which its flux
fell below a specified threshold.  This method is a compromise between
elimination of polluting starlight and retention of usable data.  To
construct the mask, the model PSF was averaged radially to a 1-d function,
scaled to each star according to its magnitude, and used to determine the
radius at which the stellar flux fell below threshold.  A threshold of 2
ADU, well below the 7 ADU RMS noise in the co-added frame, was a
satisfactory compromise between starlight removal and pixel retention.
Typical PSF reference stars were masked out to a radius of 7 pixels
(16\arcsec), and the brightest unsaturated stars were masked out to a
radius of 9 pixels (21\arcsec), about 26 times larger than the average
seeing HWHM of about 0.8\arcsec.

\begin{figure*}[p]
\plotone{f2b.eps}
\\
Figure~\ref{fig:m83chanmaps} (continued)
\end{figure*}

\begin{figure*}[t]
\plotone{f2c.eps}
\\
Figure~\ref{fig:m83chanmaps} (continued)
\end{figure*}

%\clearpage

\begin{figure*}[htp]
\plotone{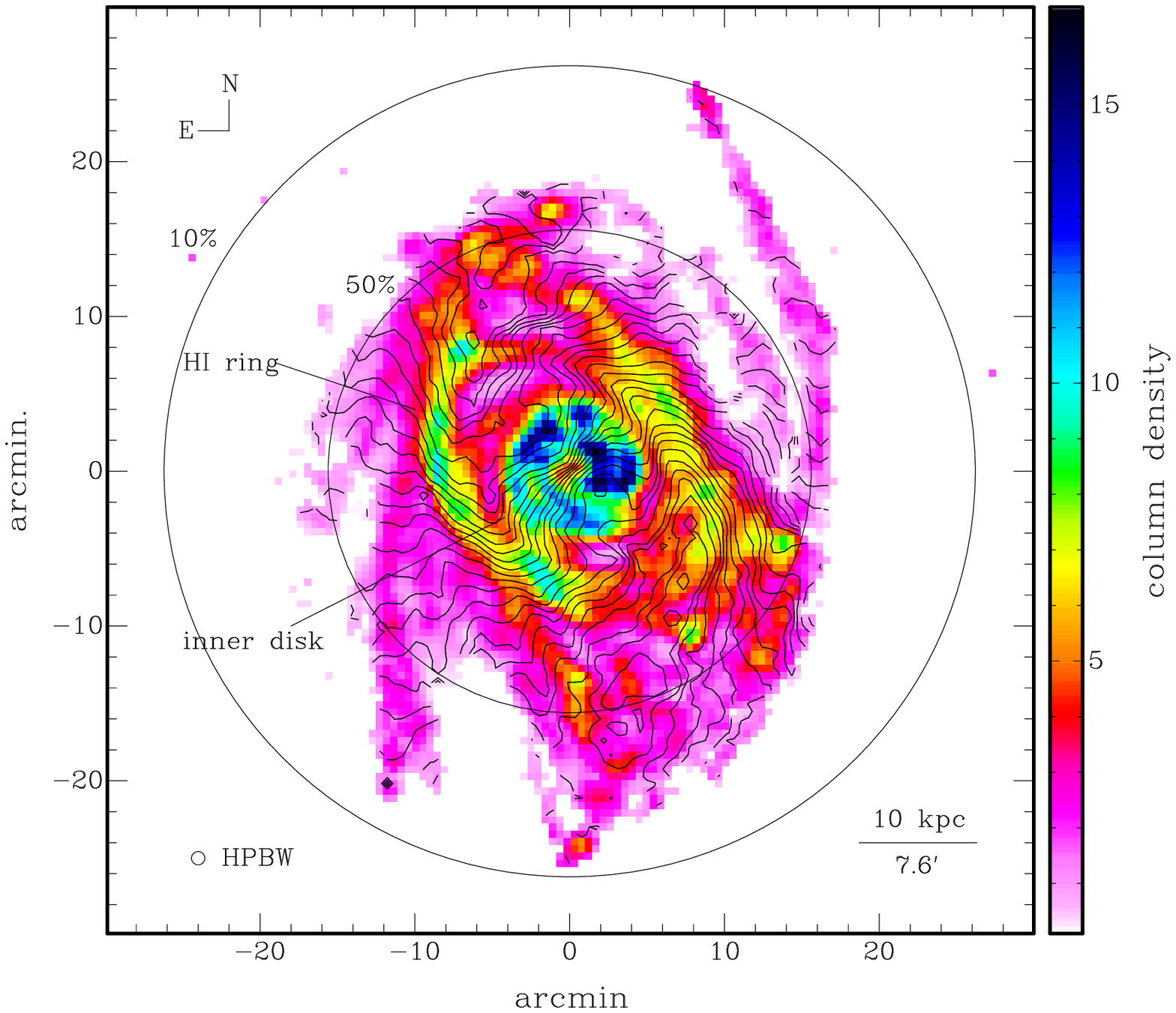}
\caption{\hi\ column density map for M 83.  The inner circle shows the FWHM
of the primary beam, while the outer circle shows the full width at 10\% of
peak sensitivity.  Contours show the mean \hi\ velocity and range from 350
to 650 \kps\ in steps of 10 \kps\ (twice the velocity resolution).  Column
density is in units of \eez{20} \cm.  The marked features are discussed in
Section \ref{subsect:m83:hidisk:disk}.}
\label{fig:m83mom0}
\end{figure*}

Some contaminating features did not conform to the shape of the PSF and
were ignored by the above procedure.  These included background galaxies,
very bright saturated stars, diffraction spikes, and charge transfer
(``bleeding'') features, and they were masked by hand.

%%%%%%%%%%%%%%%%%%%%%%%%%%%%%%%%%%%%%%%%%%%%%%%%%%%%%%%%%%%%%%%%%%%%%%%%
% SECTION -- The Normal Galactic Disk of M 83
%%%%%%%%%%%%%%%%%%%%%%%%%%%%%%%%%%%%%%%%%%%%%%%%%%%%%%%%%%%%%%%%%%%%%%%%
\section{The Normal Galactic Disk of M 83} 
\label{subsect:m83:hidisk}

\subsection{\hi\ Moment Map Construction} 
\label{subsect:m83:hidisk:mom}

The bulk distribution and kinematics of emission-line gas are most easily
summarized by moment maps, including the total \hi\ surface brightness (0th
moment) and the intensity-weighted mean velocity field (1st moment).  To
construct the moment maps, the image cube was first smoothed spatially to a
round beam of 52\arcsec\ FWHM, roughly 1.5 times the size of the original
synthesized beam.  This smoothing is done to improve the S/N and ensure
that the faint extended components of emission peaks are included in the
summation.  Pixels with flux greater than 4-$\sigma$ were included,
as were immediate neighbor pixels in both space and velocity.

\begin{figure*}[htp]
\plotone{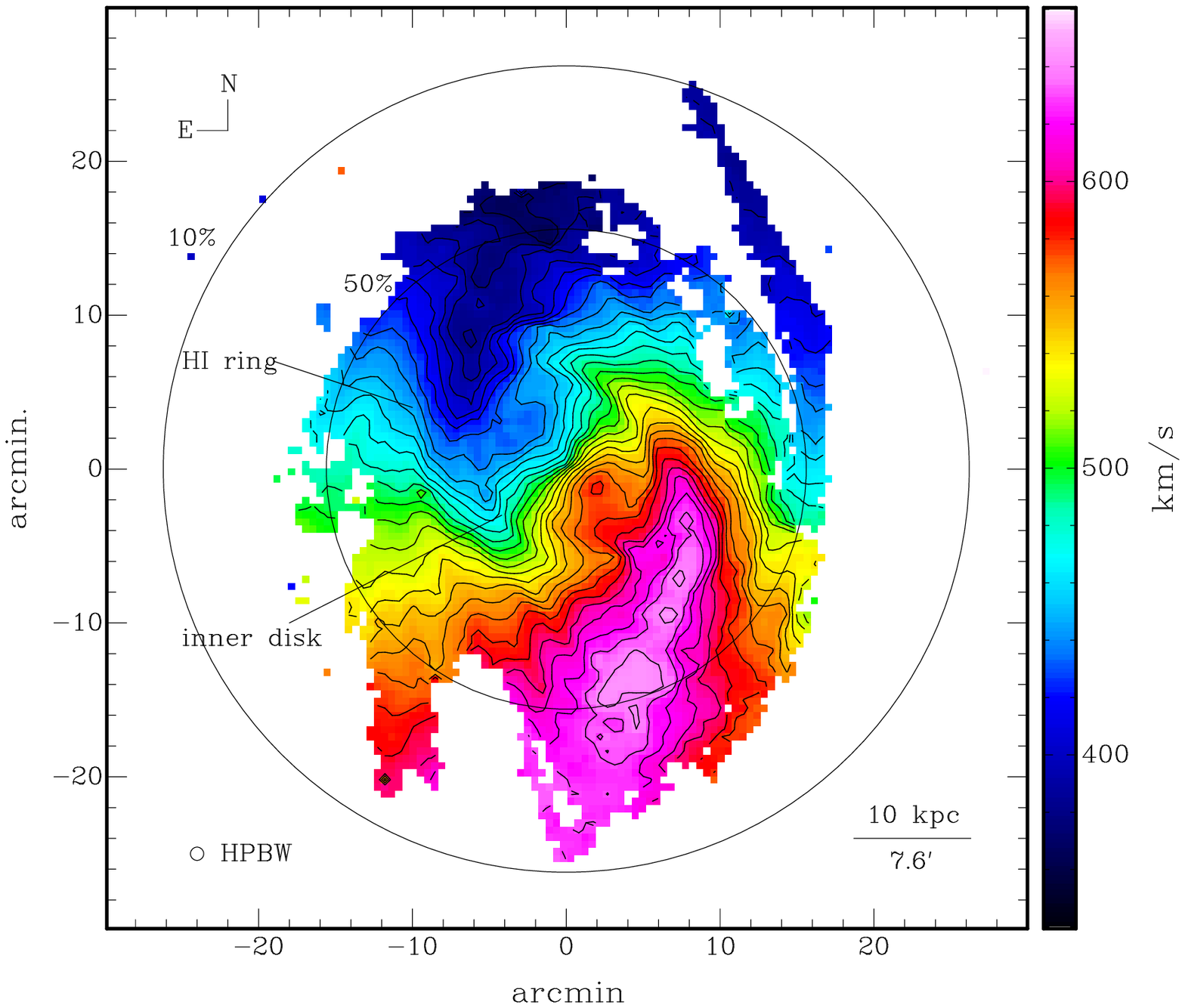}
\caption{Intensity-weighted mean velocity map for M 83.  Notations and
contours are the same as in Figure \ref{fig:m83mom0}, and velocities are
heliocentric.}
\label{fig:m83mom1}
\end{figure*}

The moment maps are shown in Figures \ref{fig:m83mom0} and
\ref{fig:m83mom1}.  The primary beam attenuation correction was applied to
only the moment 0 map (which is shown in column density units) after
integrating in velocity.  Throughout this work, all \hi\ intensity maps
displaying either integrated \hi\ emission or \hi\ column density have been
corrected for the primary beam sensitivity.

The total \hi\ mass determined from the \hi\ surface brightness map is
4.7\eex{9} \msun. This is lower than the distance-corrected value of
6.1\eex{9} obtained by \citet{HuchtBohn81}, who mapped the emission out to
about twice the radius as the present study.  Their single-dish scans were
impervious to zero-spacing issues of interferometry and recovered all of
the emission in the beam.  Our result is therefore in reasonable agreement
with their value.

\subsection{\hi\ Disk Features} 
\label{subsect:m83:hidisk:disk}

The total \hi\ map (Figure \ref{fig:m83mom0}) is a good tracer of spatial
morphology, and it shows a number of notable features.  The inner disk of M
83 contains a minimum in the \hi\ approximately 1\arcmin\ (1.3 kpc) in
diameter, with a measured column density of 4\eex{20} \cm.  This feature is
surrounded by the inner gaseous disk, which extends to a radius of
6\arcmin\ (8 kpc) and ranges in column density from a minimum of
7\eex{20} \cm\ to a (probably optically thick) maximum of 1.7\eex{21} \cm.
The inner disk shows a hint of spiral structure, with an irregular pattern
of bright spots, and an aspect ratio close to unity, as one would expect
for a nearly face-on galaxy.  The small-scale structure is unresolved in
our map, but it agrees qualitatively with the results of
\citet{TilanusAllenM83}, who studied M 83 at higher spatial resolution.

Beyond 5\arcmin, the \hi\ brightness falls off abruptly, with $\nh \lesssim
4\eex{20}$ \cm\ in the region immediately around the inner disk.  In the
regions NE and SW of the inner disk, the column density becomes very low,
approaching 5\eex{19} \cm.  Further out, there is evidence of a bright ring
of \hi\ at a radius of 10\arcmin\ (13 kpc).  The ring has a peak column
density of 1\eex{21} \cm, similar to the inner disk, and the projected
thickness varies between 60\arcsec\ and 80\arcsec.  With the smoothed beam
size (52\arcsec) this scale is unresolved, but the width can be no more
than 1\arcmin\ (1.3 kpc) after removal of the instrumental profile.  The
emission in the ring appears clumpy, with the brightest region in an arc on
the west side of the galaxy.  In addition, the aspect ratio of the ring is
quite different from that of the inner disk, with semi-major and semi-minor
axes of 10\arcmin\ and 5\arcmin\, respectively.  The position angle is
close to that of the inner disk (see below).

Outside of the ring, the \hi\ spatial distribution becomes irregular, with
arms and streams at various projected orientations.  One arm
appears just beyond the ring on the NW side, and spans about 25\arcmin\ (33
kpc) to a point 16\arcmin\ (21 kpc) NE of the disk center.  There is a
corresponding arm on the SE to S side of the galaxy, at a similar
projected distance.  Two notable features are the arms further west of
this, which extend from a common point 14\arcmin\ (18 kpc) west of the
center and extend toward the north.  The longer of these arms reaches a
projected galactic radius of 26\arcmin\ (34 kpc) and is clearly detached
from the rest of the disk.  Both NW arms appear to be mirrored in the SE,
and similar mirroring can be seen between other features and clumps in the
extended \hi.  Since the primary antenna sensitivity falls to 10\% of the
central value for the most extended arms, these features could be even more
extended than observed.  It is clear that the arms are distinct from the
main \hi\ disk, down to our detection limit.  The symmetrical nature and
varying orientations of the outer arms indicates either a warp that varies
with radius or a collection of kinematically distinct orbiting streams.

Analysis of the kinematic morphology can improve our understanding of the
complex outer disk features, and this is most thoroughly done by combining
the velocity field map, the individual channel maps, and the data cube as a
whole.  The velocity map is presented in both Figure \ref{fig:m83mom0} (as
contours) and Figure \ref{fig:m83mom1} (as an image map and contours).
In both figures, the inner disk shows a typical differential rotation
pattern (a so-called ``spider diagram''), with a central systemic value of
$\sim 500$ \kps.  To estimate the systemic velocity and orientation, we fit
the central 5\arcmin\ with a Brandt rotation curve using the {\tt GAL} task
in {\tt AIPS} \citep{AIPSCookbook}.  We find a systemic velocity of 
$v_{\rm hel} = 513 \pm 2$ \kps, where the error is estimated from the
spread of best-fit values at different radii, and using the optical
definition
$v_{\rm hel,opt} = c (\lambda_{\rm obs}-\lambda_{\rm ref}) / \lambda_{\rm
ref}$.  The best-fit position angle is $225\arcdeg \pm 2\arcdeg$ (measured
to the receding side from north to east), and the best-fit inclination is
$i = 23\arcdeg \pm 2\arcdeg$.

Our estimates of the inner \hi\ disk parameters are consistent with
previous results.  Previous values for the systemic velocity range from 505
\kps\ determined by \citet{Comte81} through \halpha\ interferometry to 516
\kps\ quoted by \citet{RC3} and determined by \hi\ observations.  The
HIPASS Bright Galaxy Catalog reports a value of $513 \pm 2$ \kps\
\citep{HIPASS04}.  We adopt 513 \kps\ as the systemic velocity for the
remainder of this work.  Our estimates of the orientation and inclination
are consistent with the those quoted by \citet{TilanusAllenM83}, who have
data with higher spatial resolution.  Therefore we adopt $i = 22\arcdeg$
for our inclination, and note that the maximum rotation rate of the
inner disk is $v_{\rm rot,max} = (v_{\rm hel,max}-v_{\rm sys}) / \sin{i} =
200$ \kps.

Outside of the inner disk, a discontinuity develops in the rotational
velocity.  This is most clearly seen in Figure \ref{fig:m83mom0}, where the
velocity contours bend abruptly into a warp, and in the channel maps for
440 \kps $< \vhel <$ 540 \kps\ (see Figure \ref{fig:m83chanmaps}).  On the
W side of the galaxy, the warp correlates well with the \hi\ ring in
projection, but on the E side the warp peaks inside of the ring.  The
isovelocity contours are shifted by 2\arcmin\ counter-clockwise from where
they would be under ordinary differential galactic rotation.  Beyond the
ring, the isovelocity contours switch direction again, and they no longer
reflect a flat rotation curve.  In addition, the aspect ratio of the disk
increases in this region, indicating that the axis of rotation is becoming
more inclined to our line of sight as the radius increases.  The highest
radial velocities are seen near the very edge of the detected \hi\
distribution, at a distance of 8\arcmin\ (11 kpc) N and S of the galactic
center.

It is apparent from this analysis that the inner disk, the ring, and the
outer arms are kinematically distinct features.  The ring and outer arms
could be considered anomalous features themselves, however we wish to
search for anomalous-velocity emission superposed on the bulk \hi\
emission, and for this purpose we assume the outer arms are part of an
extended disk of varying inclination.  

\subsection{Large-Scale Optical Features}

\begin{figure*}
\plotone{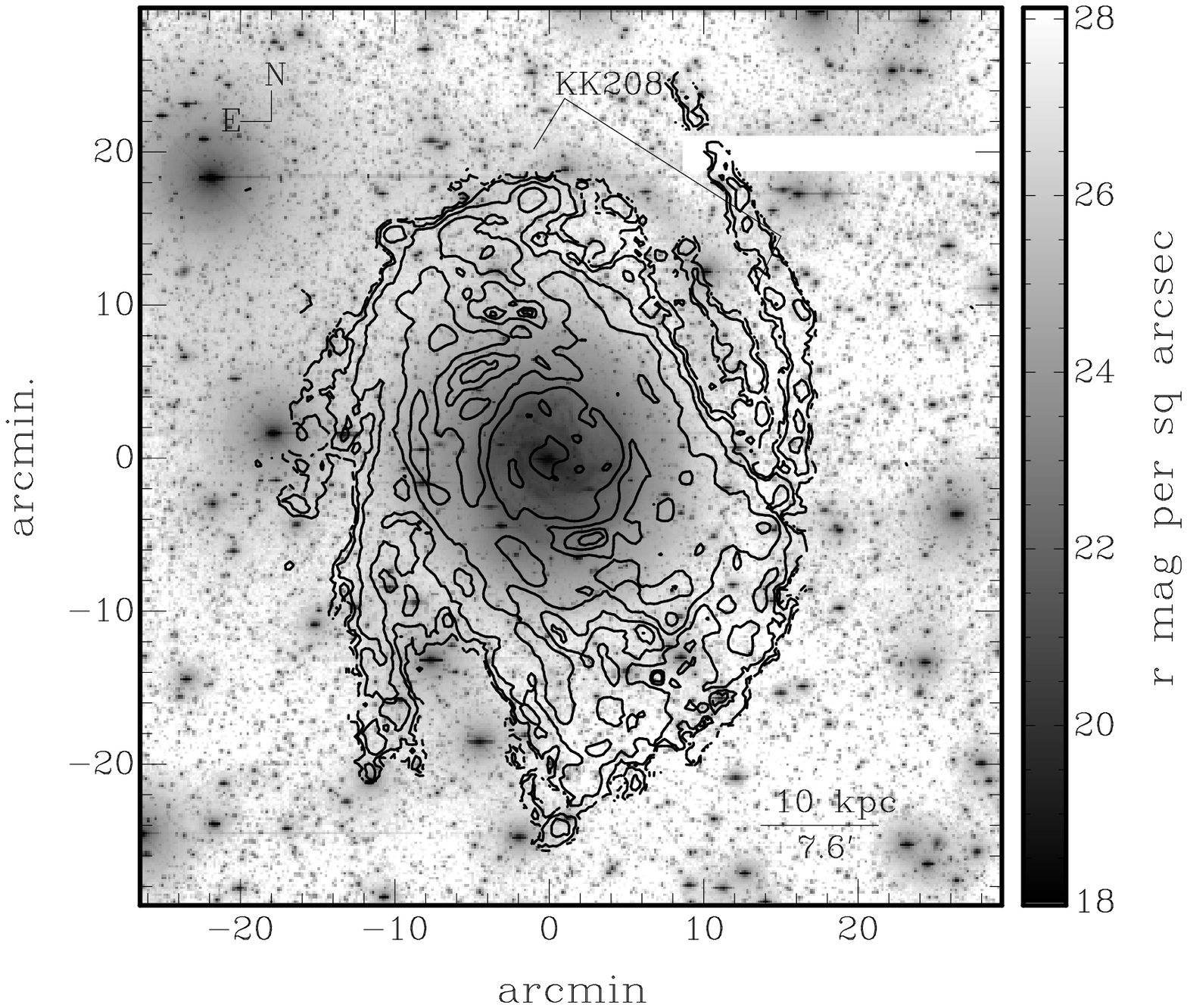}
\caption{Optical image for M 83, displayed in linear units of $\rsloan$
magnitude per square arcsec. The contours show the column density of \hi,
starting at 5\eex{19} \cm\ and increasing by factors of 2.  Also labeled is
KK208, a faint companion galaxy or stellar stream to the northeast.}
\label{fig:m83opt}
\end{figure*}

The final optical image for M 83 shows many of the typical features of a
face-on, grand-design spiral galaxy (see Figure \ref{fig:m83opt}).
The spiral arms are seen radiating from a central bar-like structure,  and
these arms contain a number of star clusters and dust lanes. The integrated
\hi\ intensity traces the spiral arms closely.  The spiral structure is
apparent only out to a radius of about 4.5\arcmin\ (6 kpc) in the north and
5.5\arcmin\ (7 kpc) in the south, at which point the surface brightness has
dropped to 22.5 \rsloan\ magnitudes per square arcsec.  This is the same
radius at which the inner \hi\ disk falls off abruptly and at which the
warp in the velocity field appears.

Beyond the inner optical disk, there is an extended optical envelope devoid
of any structure.  This envelope reaches a surface brightness of 26
\rsloan\ mag per square arcsec at a radius of about 10\arcmin\ (13 kpc),
falling to undetectable levels at the location of the \hi\ ring.  In
addition, the aspect ratio of the extended light, appearing stretched
northeast to southwest, is different from that of the inner disk, which is
nearly circular.

An arc-like feature can be seen projected 20\arcmin\ (26 kpc) northwest of
center, extending 12\arcmin\ (16 kpc) northeast to southwest.  This feature
is barely visible in the Palomar Sky Survey plate covering this region, and
it has been cataloged as a dwarf galaxy by
\citet{KarachentsevaKarachentsev98}, who designate it KK208.  The surface
brightness across the feature varies from 25--26 \rsloan\ mag per square
arcsec, although there are three bright stars along the arc that complicate
masking and flux measurements.  There is no measurable \hi\ anywhere along
the feature down to a 3-$\sigma$ column density limit of 1.5\eex{19}
\cm.  The arc lies between two outer \hi\ arms, and it has a shape similar
to those arms.

%%%%%%%%%%%%%%%%%%%%%%%%%%%%%%%%%%%%%%%%%%%%%%%%%%%%%%%%%%%%%%%%%%%%%%%%
% SECTION -- Anomalous-Velocity HI: Extended Emission
%%%%%%%%%%%%%%%%%%%%%%%%%%%%%%%%%%%%%%%%%%%%%%%%%%%%%%%%%%%%%%%%%%%%%%%%
\section{Anomalous-Velocity \hi: Extended Emission} 
\label{sect:m83:hiavext}

\subsection{\hi\ Disk Modeling} 
\label{subsect:m83:hiavext:model}

To search for anomalous-velocity material, one must first establish the
criteria that make such material ``anomalous''.  The difficulty of this can
be appreciated by considering the ensemble of Galactic HVCs.  A
commonly-used definition limits membership to complexes with velocities
deviating by greater than 100 \kps\ (or sometimes 90 \kps) relative to the
Local Standard of Rest (LSR).  This definition was intended to separate the
HVCs from the so-called Intermediate Velocity Clouds (IVCs) and gas in the
plane.  However, such a definition does not take into account differential
galactic rotation, an effect which can vary the observed velocity of a
cloud by as much as 100 \kps, depending on the line of sight direction
\citep{Wakker91a}.  To account for this effect, some authors have used the
velocity relative to the so-called Galactic Standard of Rest (GSR), which
is a point at the location of the Sun that does not participate in galactic
rotation. This definition works well for gas at high $|z|$. However, it
over-corrects the velocity of any cloud that has a rotational component of
motion, such as may be found closer to the plane.  \citet{Wakker91a}
introduced the concept of ``deviation velocity,'' which is the difference
between the observed LSR velocity of the cloud and the maximum velocity
expected along that line of sight due to differential rotation, assuming a
simple model for the gaseous Galactic disk.  At low latitudes in the Milky
Way this varies greatly, but at latitudes above $\sim 40\arcdeg$ the
velocity limit is around 60 \kps.

For external spiral galaxies, and especially face-on galaxies, the
situation is simplified in that the bulk disk velocity is easily measured
along any line of sight.  Material at velocities different from this,
assuming some velocity threshold or kinematic model for the disk, would be
kinematically distinct from the cold \hi\ disk and therefore anomalous.
Hereafter we define anomalous-velocity material to be any \hi\ emission
that does not participate in the normal differential rotation of the bulk
of \hi\ along that line of sight.  Implicit in the definition is the
inclusion of spatially isolated \hi\ clumps, since the disk rotation is
undefined in regions where the disk does not exist.  Therefore an
anomalous-velocity clump (hereafter AVC or AV gas) can be anomalous in
velocity or position.

As low-velocity AV gas could overlap and merge with the disk velocity, it
is useful to remove any \hi\ disk emission.  Additionally, it is important
to ignore this emission in a non-interactive source detection scheme.  We
modeled the disk of M 83 by fitting a Gaussian to the spectrum along every
pixel of the cleaned image cube.  The fitting was performed only along
lines of sight where the peak exceeded 3.5-$\sigma$ (3.3 mJy/beam), and
only included channels whose flux exceeded 1.7-$\sigma$ (1.6 mJy/beam).
These criteria were applied to ensure that faint, velocity-broadened
emission would not contaminate the thin disk fit.  

Disk fitting and removal failed at warp discontinuities, where the \hi\
disk has multiple velocity components across a scale of about 1\arcmin.  It
is not clear whether this spatial overlap of kinematically distinct disk
components is a real feature or an effect of beam smearing, but in either
case it poses a difficulty for effective disk removal.  It is clear that
the warp features are not AV material in our definition, since they define
the disk velocity.  We therefore attempted to remove this emission by
refitting the original data cube with multiple Gaussian components.  For
lines-of-sight where the residual flux matched the above single-Gaussian
criteria (i.e., peak above 3.5-$\sigma$, 3 or more adjacent pixels with
flux above 1.7-$\sigma$), a two-component Gaussian was fit.  The fitting
threshold was kept high so that true AV gas would escape the fit.

To estimate the quality of the disk model, we examined the reduced $\chi^2$
and covariance matrices of all the fits performed.  The fits that included
only a few data points were obviously poorly constrained, but inspection of
the residual data cube showed that even these fits were reasonable.  There
were several lines-of-sight that had noticeably bad fits, but these were
generally isolated pixels whose neighbors had similar parameters to
each other.  To remove the outliers, we performed a median box filtering on
the fitting parameter images, using a kernel width of 3 pixels.  
The disk model and residual cube were reproduced from the adjusted fitting
parameters, and inspection showed that the poor fits were removed and the
disk subtraction was improved.  

\begin{figure*}[p]
\plotone{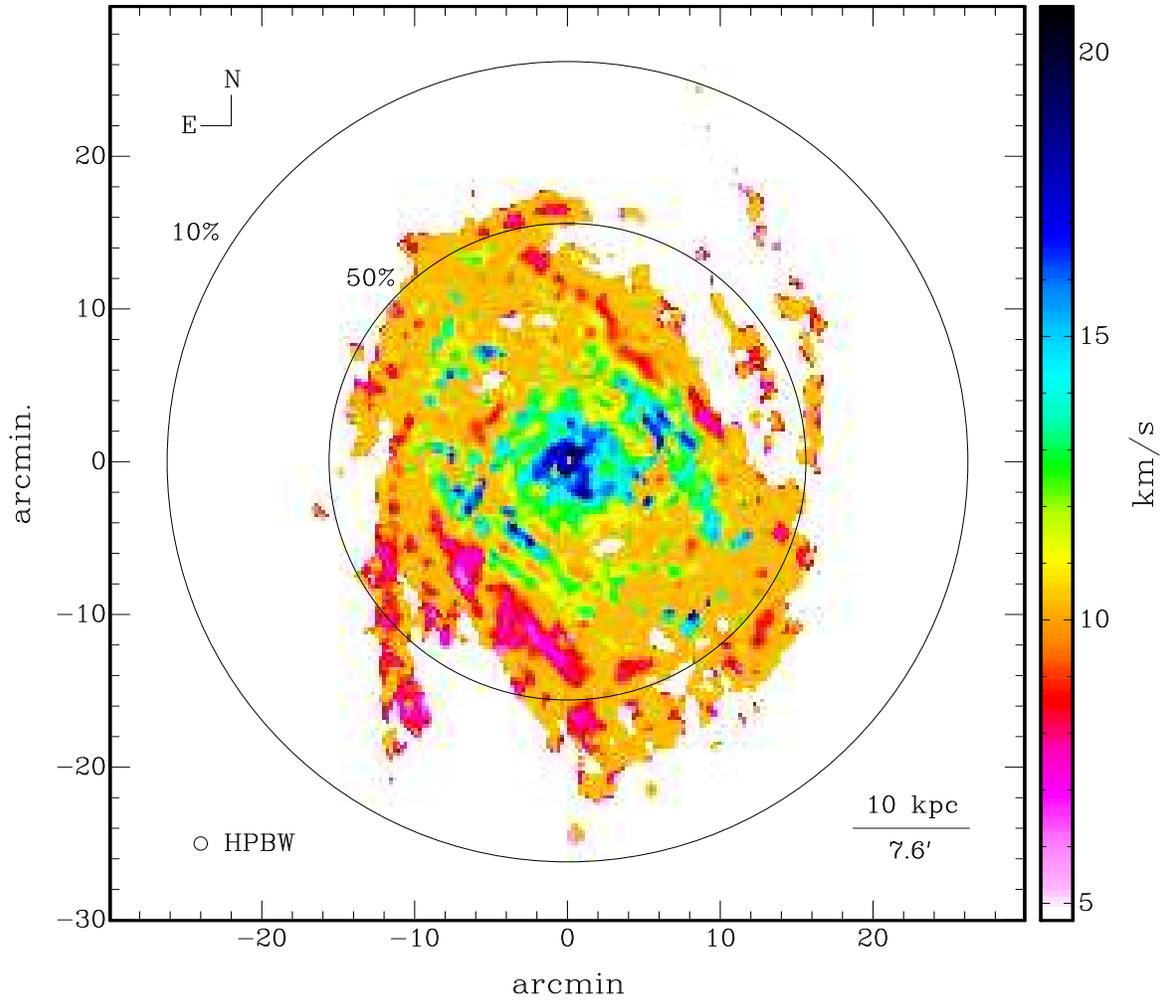}
\caption{Line-of-sight \hi\ velocity dispersion for M 83, obtained by
fitting a Gaussian model to the disk emission.  The fit was performed along
every line of sight (pixel), but the values shown have been median box
filtered to remove poor fits.  The velocity dispersion is 20 \kps\ in the
center, falling to less than 10 \kps\ in the outer regions.}
\label{fig:m83gparsig}
\end{figure*}

The fitted line-of-sight velocity dispersion varies across the galaxy,
as can be seen in Figure \ref{fig:m83gparsig}.  It reaches a maximum of 20
\kps\ in the center, averages about 15 \kps\ in the central \hi\ disk and
ring, and falls to 10 \kps\ or less in the extended disk and arms. 

\subsection{Extended, Disk-Like AV Gas} 
\label{subsect:m83:hiavext:beard}

\begin{figure*}[p]
\plotone{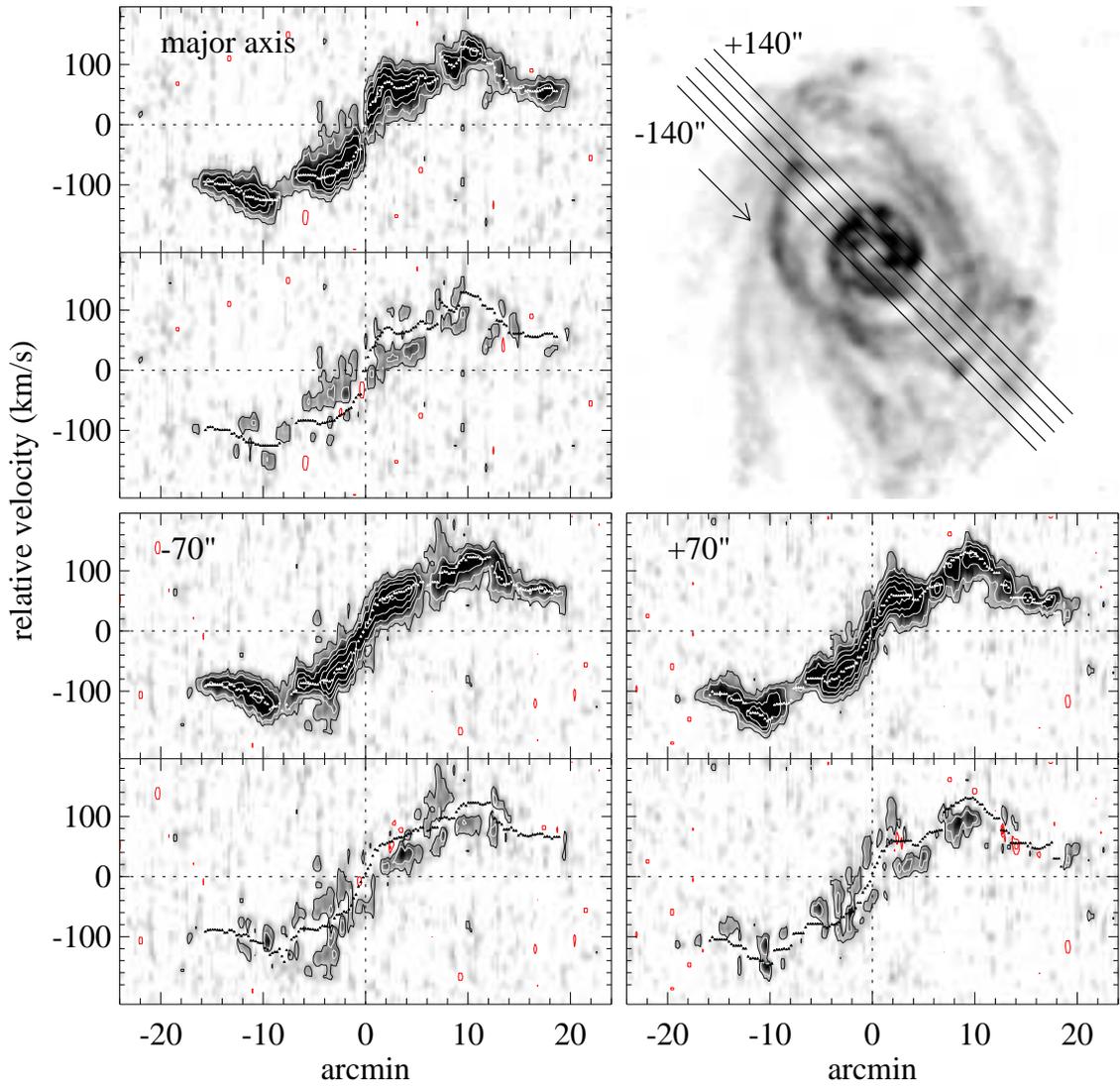}
\caption{Position-velocity (\pv) diagrams for M 83, taken parallel to the
major axis at separations of 70\arcsec, as plotted on the \hi\ emission map
(upper right) with the slice direction indicated by the arrow.  Each slice,
boxcar smoothed in velocity by 3 channels, is shown with the disk emission
included (top) and subtracted (bottom).  The rotation curve is traced by
the white and black points.  Grayscale ranges from 0 to 6 mJy/beam. Black
and white contours begin at $+2$-$\sigma$ and increase by factors of two.
Red contours begin at $-2$-$\sigma$ and decrease by factors of two.
Residual emission can be seen in all the slices, especially between the
systemic and rotational velocity along the major axis.}
\label{fig:m83slices}
\end{figure*}

Removal of the \hi\ disk results in a residual cube with significant \hi\
emission remaining.  For ease of display, we created position-velocity
(\pv) plots, or slices through velocity along a line of specified location,
length, and position angle.  The slices were taken parallel to the major
axis at a spacing of 70\arcsec\ (twice the beamwidth).  The value of a
given pixel in the \pv\ image equals the distance-weighted average of the
nearest four pixels in the channel.  A sample of slices is shown in Figure
\ref{fig:m83slices} for both the full data cube and the residual.  

\begin{deluxetable}{ccccccc}
\tabletypesize{\normalsize}
\tablewidth{0pt}
\tablecaption{Extended AV Gas: Dynamical Properties
     \label{tab:m83beard}}
\tablehead{
\colhead{Component} &
\colhead{$F$} &
\colhead{$\mhi$} &
\colhead{KE} \\
\colhead{} &
\colhead{(Jy \kps)} &
\colhead{(\eez{7} \msun)} &
\colhead{(\eez{53} erg)} 
}
\startdata
Full residual cube   &    203.0 &    97.0 &    21.0 \\
Inner disk           & \phn17.6 & \phn8.4 &    16.0 \\
Inner disk, low-$v$  & \phn11.8 & \phn5.6 & \phn8.7 \\
\enddata
\end{deluxetable}

Comparison of the slices with and without the disk emission clearly shows
an emission component not well modeled by differential galactic rotation.
This is especially evident along the major axis.  The anomalous component
is spatially extended and clumpy, with deviation velocities generally
$\pm40$--50 \kps\ near the emission peak along any line of sight.  The
emission generally appears brighter on one velocity wing of the disk \hi,
toward the systemic galactic velocity\footnote{Hereafter, velocities on the
systemic side of the disk velocity will be referred to as ``low relative
velocity'', while velocities on the opposite side of the disk velocity will
be referred to as ``high relative velocity''.  This designation places the
velocities in the reference frame of the rotating galaxy.}.  Some of the
anomalous emission occurs in the ``forbidden'' velocity region, which is
the \pv\ quadrant on the opposite side of the systemic velocity from the
disk.  One interpretation is that material at velocities between the disk
and galactic systemic would be rotating more slowly than the bulk of
material in the galaxy.  Material in the forbidden \pv\ region would be
moving in the opposite sense of the bulk rotation along that line of sight,
i.e., counter-rotating.  These interpretations assume the disk and
anomalous material have similar spatial distribution and rotation axis
orientation.  In a face-on galaxy like M 83, material in the forbidden
region is likely to be in vertical motion.

\begin{figure*}[t]
\plotone{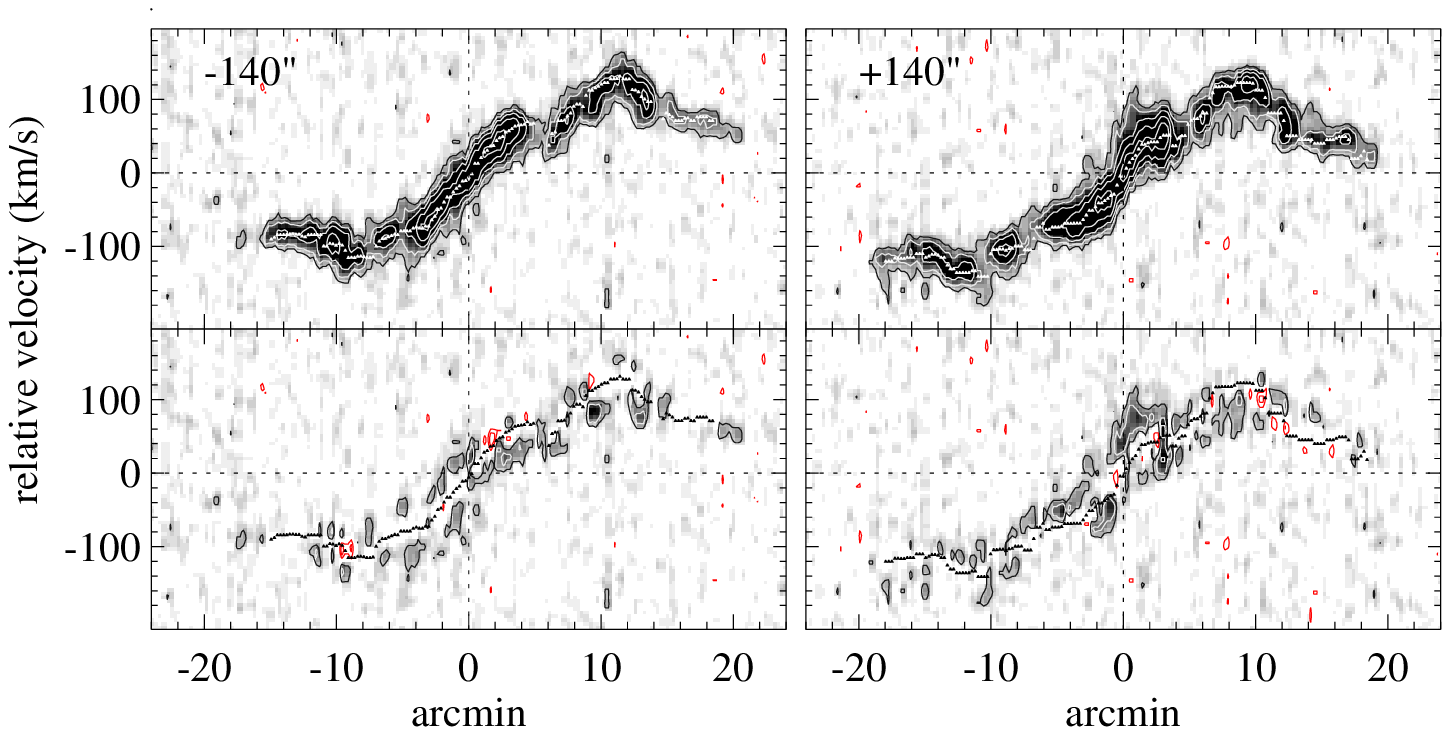}
\\
Figure~\ref{fig:m83slices} (continued)
\end{figure*}

\begin{figure*}[p]
\plottwo{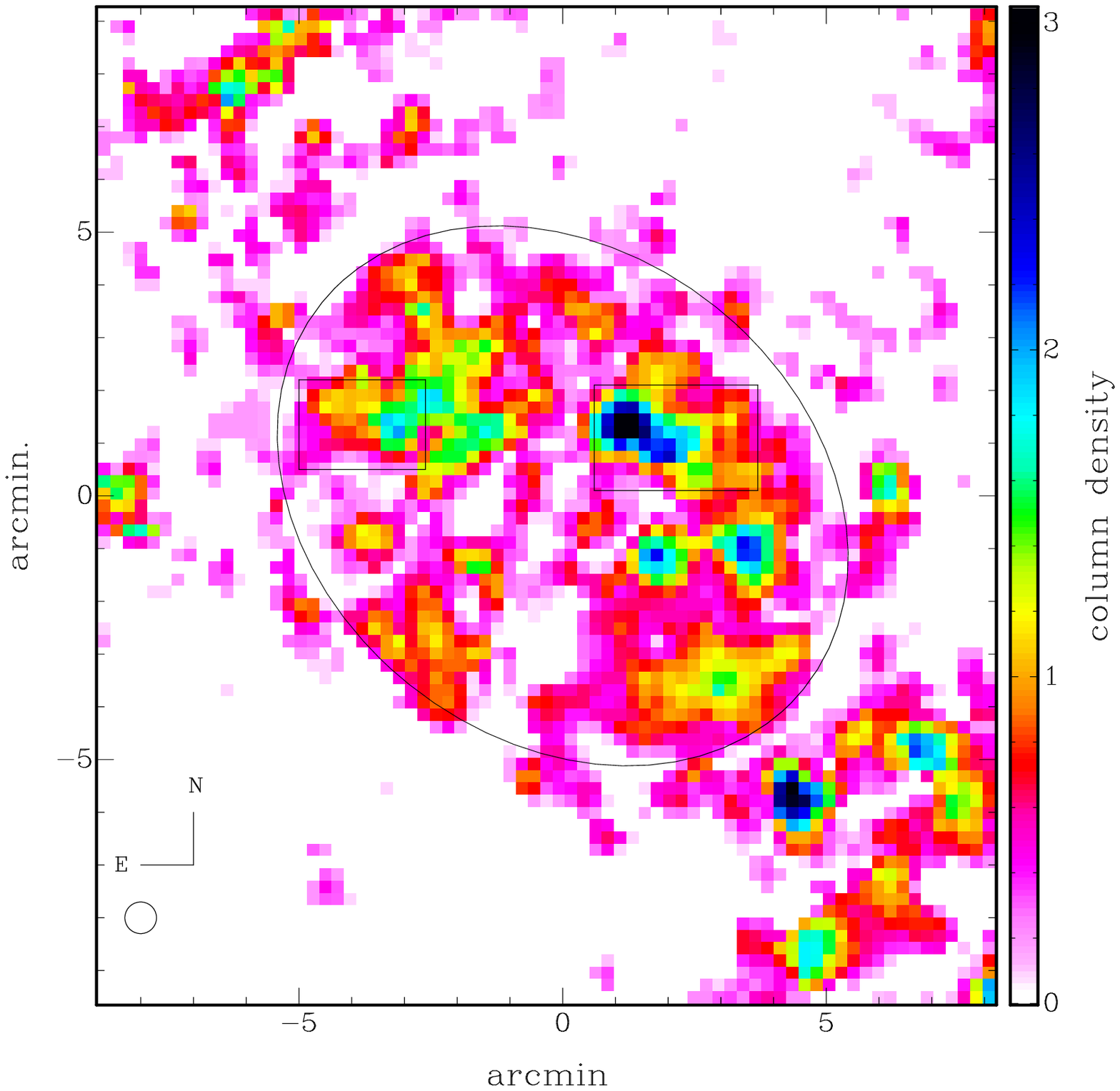}{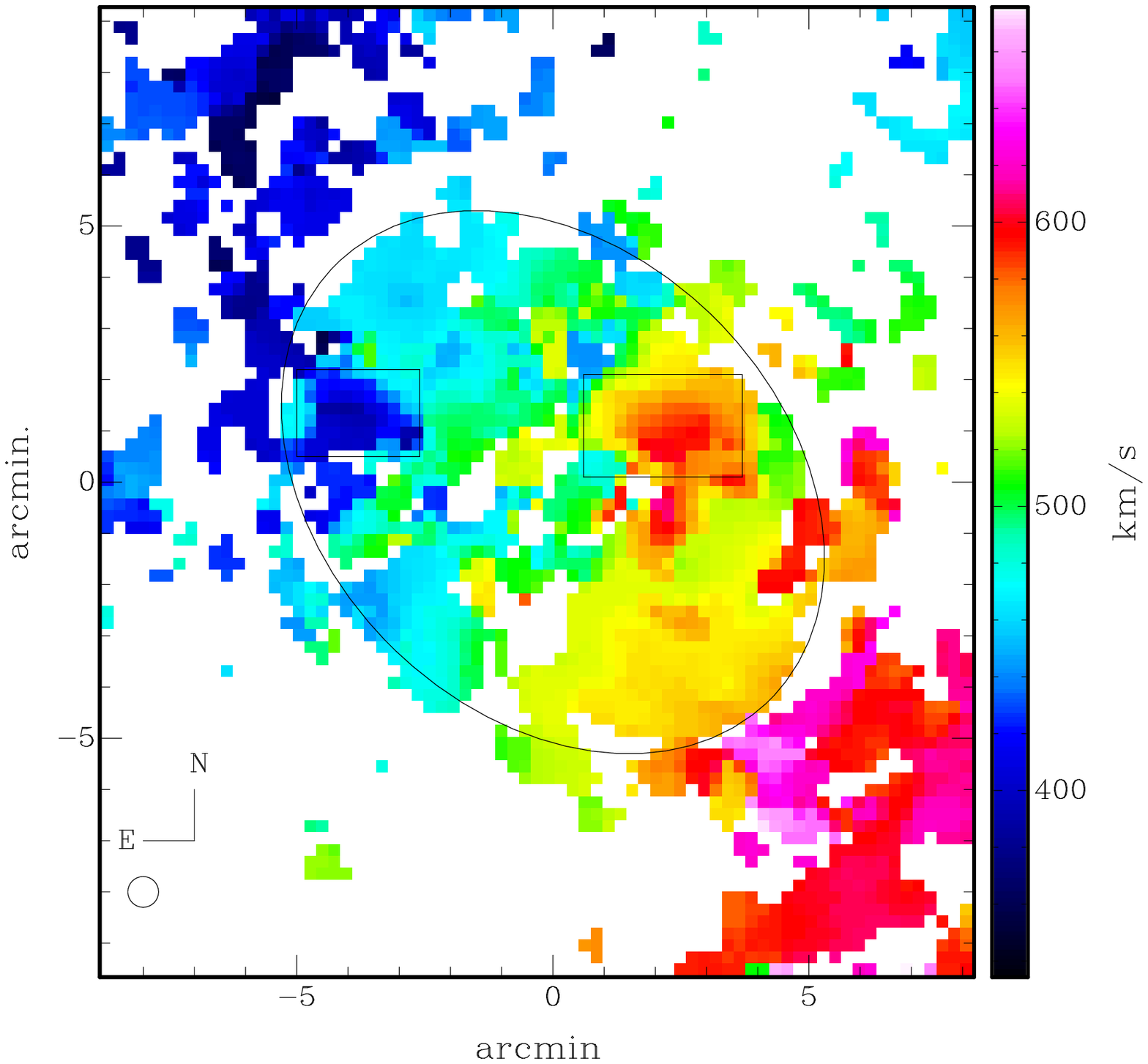}
\caption{Maps of the AV emission in M 83, showing ({\it left\/}) column
density (in units of \eez{20} \cm) and ({\it right\/}) mean velocity.  The
ellipse indicates the inner, well-modeled portion of the cold \hi\ disk,
and the rectangles identify discreet high-relative-velocity features, as
described in the text.  The anomalous gas appears disk-like, rotating about
40--50 \kps\ more slowly in projection than the cold \hi\ disk.}
\label{fig:m83beard1}
\end{figure*}

To quantify the amount of AV gas in M 83, we constructed column density and
mean velocity maps from the residual cube (see Figure \ref{fig:m83beard1}).
The dynamical properties of the AV \hi\ were determined from these maps and
are summarized in Table \ref{tab:m83beard}.  The mass of AV \hi\ in the
residual cube is 9.7\eex{8} \msun, corresponding to 20\% of the total \hi.
It was apparent that poor subtraction of the \hi\ ring/warp and outer \hi\
arms contaminated the residual cube, therefore we manually defined a region
containing only anomalous emission projected on the central disk.  The
region used was an ellipse centered on the galaxy with a semi-major axis of
6.0\arcmin (radius of 8 kpc), a semi-minor axis of 4.5\arcmin, and a
position angle of 45\arcdeg\, as shown in Figures \ref{fig:m83beard1}.
This area contains the most dynamically homogeneous disk gas and produced
reasonable model parameters.  The total \hi\ mass in this region, including
the cold disk, is 1.02\eex{9} \msun, of which 8.4\eex{7} \msun\ or 8.2\% is
present in the anomalous component (``inner disk'' in Table
\ref{tab:m83beard}).  The low-relative-velocity emission contributes
5.6\eex{7} \msun\ or 5.5\% of the total inner disk \hi.

The velocity field shows two high-velocity spots to the east and west of
the galaxy center, identified by rectangles in Figure \ref{fig:m83beard1}.
These features are also seen on the high relative velocity side of the
\pv\ slices, including one structure $+1$\arcmin\ along the $+140$\arcsec\
\pv\ diagram in Figure \ref{fig:m83slices}, at velocities between $+$50
and $+$100 \kps.  In a simple model of disk rotation, these clumps would
be rotating more quickly than the bulk of the \hi.  They are discrete in
projection compared to the low relative velocity component described
above, and will be discussed in detail in Section \ref{sect:m83:hiavcs}
(as AVCs 1 and 2).

\begin{figure*}[p]
\plottwo{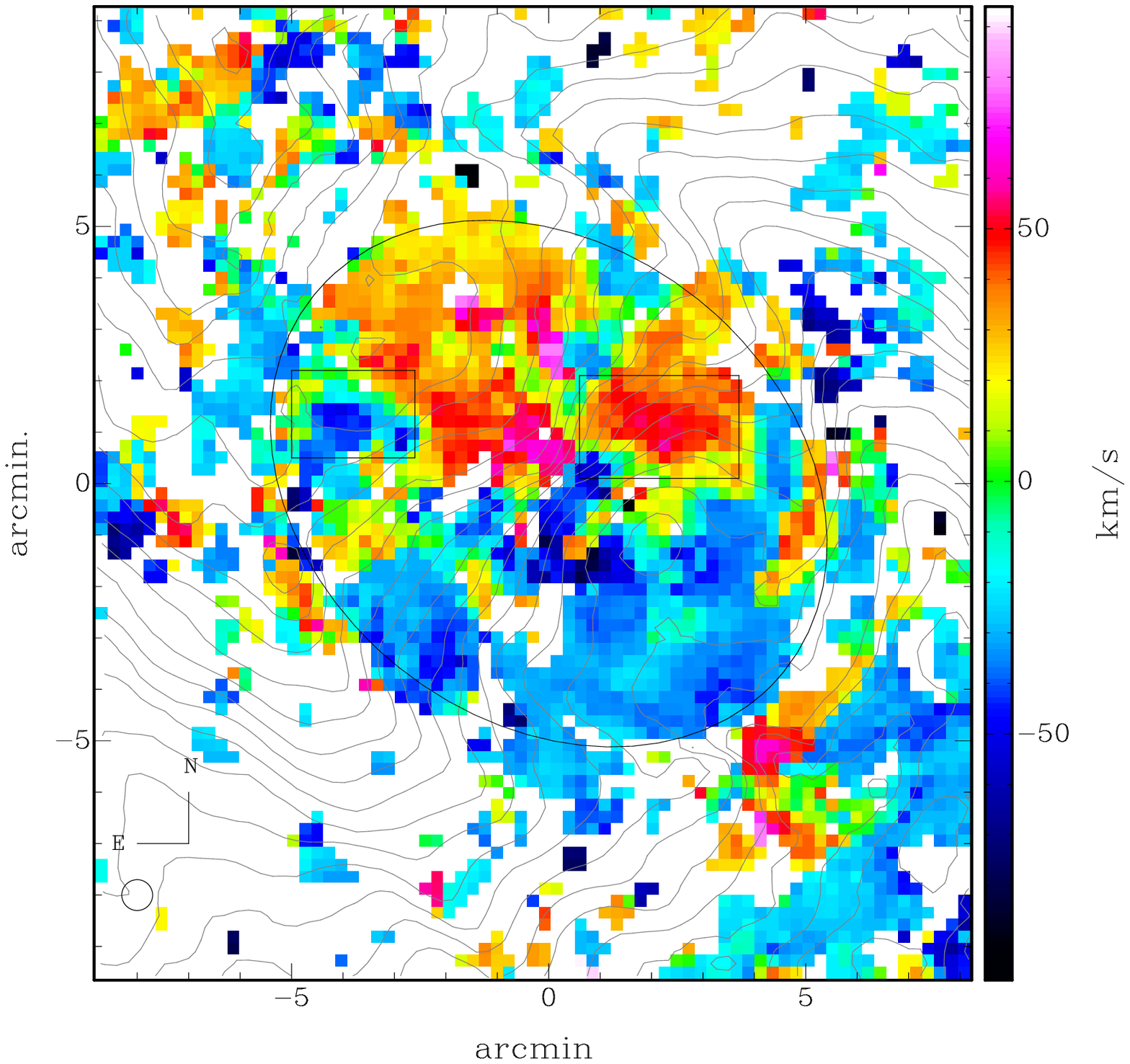}{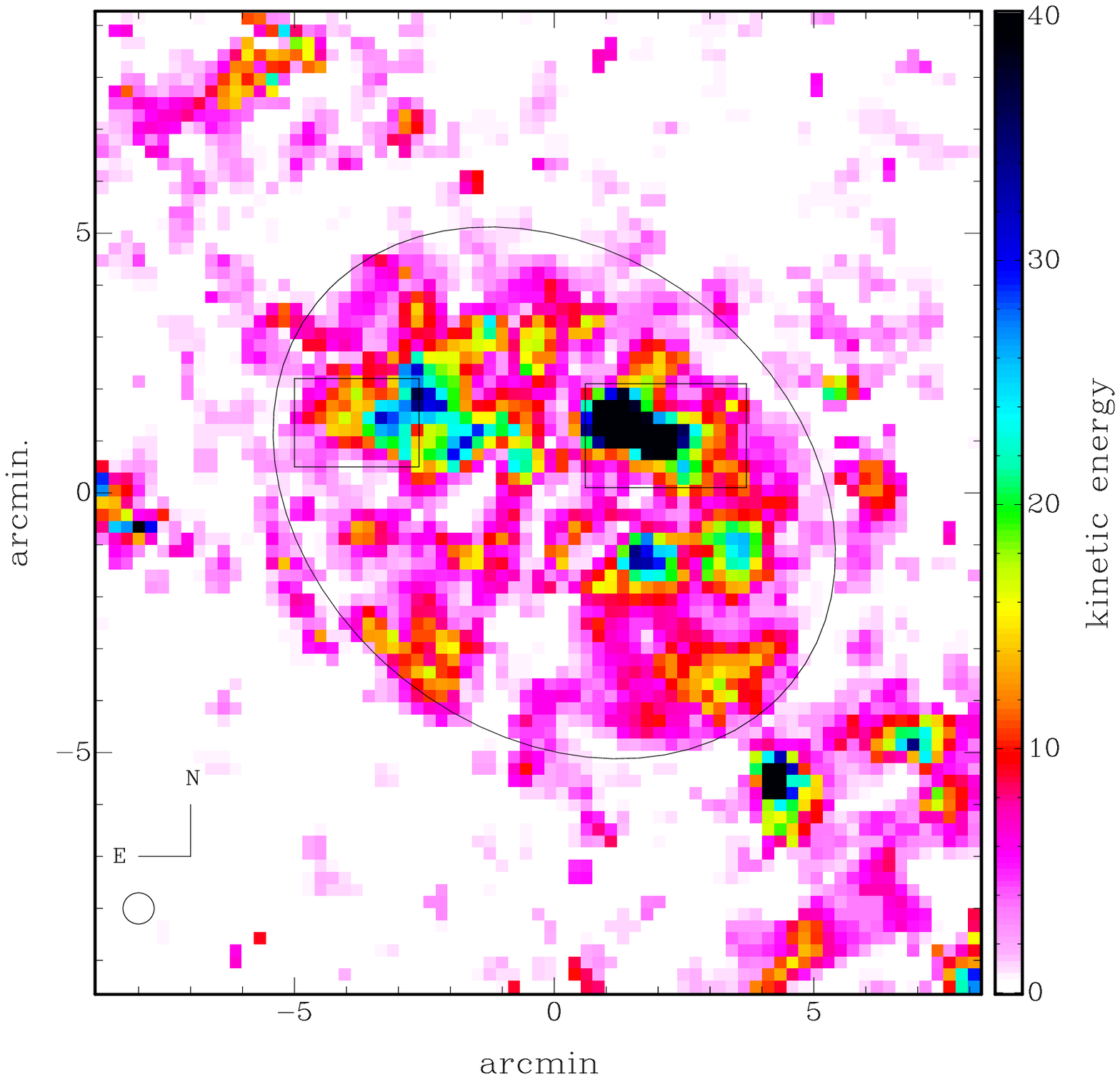}
\caption{Maps of the AV emission in M 83, showing ({\it left\/}) deviation
velocity and ({\it right\/}) kinetic energy per $14\arcsec \times
14\arcsec$ pixel (in units of \eez{50} erg).  The contours show the mean
\hi\ disk velocity in steps of 10 \kps\ for reference, and the other
notations are as described in Figure \ref{fig:m83beard1}.  The anomalous
gas appears disk-like, rotating about 40--50 \kps\ more slowly in
projection than the cold \hi\ disk, but with about the same rotation axis.
Most of the kinetic energy is contained in the feature 2\arcmin\ northwest
of center.}
\label{fig:m83beard2}
\end{figure*}

Aside from these features, the motion and morphology of the residual
emission mimic that of the galactic disk, with the southwest side
receding and the northeast side approaching.  The line-of-sight velocity
dispersion ranges from 10--15 \kps, similar to the thin disk, although this
may be an underestimate if the wings of the anomalous emission are excluded
in the thin disk mask.  A map of deviation velocity was produced by
subtracting the velocity map of the modeled disk from that of the residual
cube (see Figure \ref{fig:m83beard2}), and it shows a rotation rate 40--50
\kps\ slower in projection.  If the inclination of this disk of AV emission
is also about 20\arcdeg, then its rotation rate is about 100--150 \kps,
compared to 200 \kps\ for the cold disk.  

We used the deviation velocities to construct a map of the kinetic energy
with respect to the cold disk (see Figure \ref{fig:m83beard2}).  Unlike
the previous analysis, this interpretation assumes all motion is
perpendicular to the galaxy plane\footnote{Lacking any knowledge of
transverse motion, we actually assumed motion fully along the line of
sight, which is offset by 24\arcdeg\ from the galactic pole direction.
Thus velocities and kinetic energies are 10\% and 20\% lower, respectively,
then what would be calculated for gas with only $z$-direction motion.}, and
it provides an upper limit to the energy apportioned to vertical bulk
motion.  The total kinetic energy so derived is 2.1\eex{54} erg, with
1.6\eex{54} erg from gas within the inner disk and 8.7\eex{53} erg of that
from the low-relative-velocity material.  The last figure excludes discrete
clumps, such as the one projected 2\arcmin\ northwest of center; these
features possess nearly half of the kinetic energy of the inner AV gas, and
are fully discussed in Section \ref{sect:m83:hiavcs}.  The 1.6\eex{54} erg
from the inner 8 kpc is equivalent to the energy of $~ 8$ supernovae
kpc$^{-2}$, although the $z$ component of velocity is not known for this
material, so this is an upper limit to the kinetic energy.

While the distribution of extended gas is well-characterized by moment maps
and \pv\ slices, faint, compact emission sources are easily overlooked.
Detection of such systems requires a quantitative, objective searching
technique that utilizes the entire set of data in a statistical way. The
next section addresses our search for faint, discrete \hi\ emission
sources.

%%%%%%%%%%%%%%%%%%%%%%%%%%%%%%%%%%%%%%%%%%%%%%%%%%%%%%%%%%%%%%%%%%%%%%%%
% SECTION -- Anomalous-Velocity HI: Discrete Emission
%%%%%%%%%%%%%%%%%%%%%%%%%%%%%%%%%%%%%%%%%%%%%%%%%%%%%%%%%%%%%%%%%%%%%%%%
\section{Anomalous-Velocity \hi: Discrete Emission} 
\label{sect:m83:hiavcs}

\subsection{Searching for AVCs} 
\label{subsect:m83:hiavcs:clumpfind}

Compared to other observational subfields of astronomy, radio synthesis
imaging suffers from a lack of robust, objective source detection software.
Identifying sources and measuring their fluxes and other parameters is
typically done by hand from moment maps and similar data products.  For
bright, point-like sources, this is not a problem, but difficulties arise
when one wishes to search statistically for emission of unknown scale and
location that may be projected on much stronger sources.

We have developed a suite of software that searches a 3-d data cube for
signal in a statistical way.  The software is described in detail in
Appendix \ref{app:snrch} and summarized here.  The data cube is smoothed
with a tunable velocity filter to emphasize sources of a particular
velocity width, following the method of \citet{Uson91}.  Peaks in this
smoothed cube are identified and used to define 3-d ``islands'' of emission
by way of contouring, using a variation of the CLUMPFIND algorithm
developed by \citet{WilliamsdeGeusBlitz94}.  The velocity smoothing allows
detection of kinematically broad but statistically significant features
that might otherwise be overlooked.  In addition, this algorithm allows a
quantitative analysis of errors with the use of simulated datasets.

To search for discrete sources, we masked the residual \hi\ cube within
$\pm 20$ \kps\ of the \hi\ disk velocity to remove poorly-subtracted
emission in the center of the disk profile.  We ran the software with a
peak threshold of 4-$\sigma$, a clump merging threshold of 2.5-$\sigma$, and
a clump extending threshold of 1-$\sigma$.  The width of the Gaussian
filtering kernel was varied from 3--9 channels (15--45 \kps).  For sources
that were detected at multiple filter widths, we used the results from the
filtering that produced the highest S/N (i.e., the kernel that was closest
in width to the unsmoothed feature).  The spatial and kinematic morphology
of the candidate clouds was analyzed interactively using a variety of
tools, including \pv\ plots, channel maps and 3-d visual rendering tools.
A two-component Gaussian fit along the velocity axis at the clump centroid
recovered the velocity and line width of the clump, as well as the line
width of any \hi\ disk emission along that line of sight.  Finally, the
flux of each clump was verified by hand using the standard radio method of
aperture photometry.  Values were consistent with the results from our
automated software.

\subsection{Detected Systems} 
\label{subsect:m83:hiavcs:avcs}

\begin{figure*}
\plotone{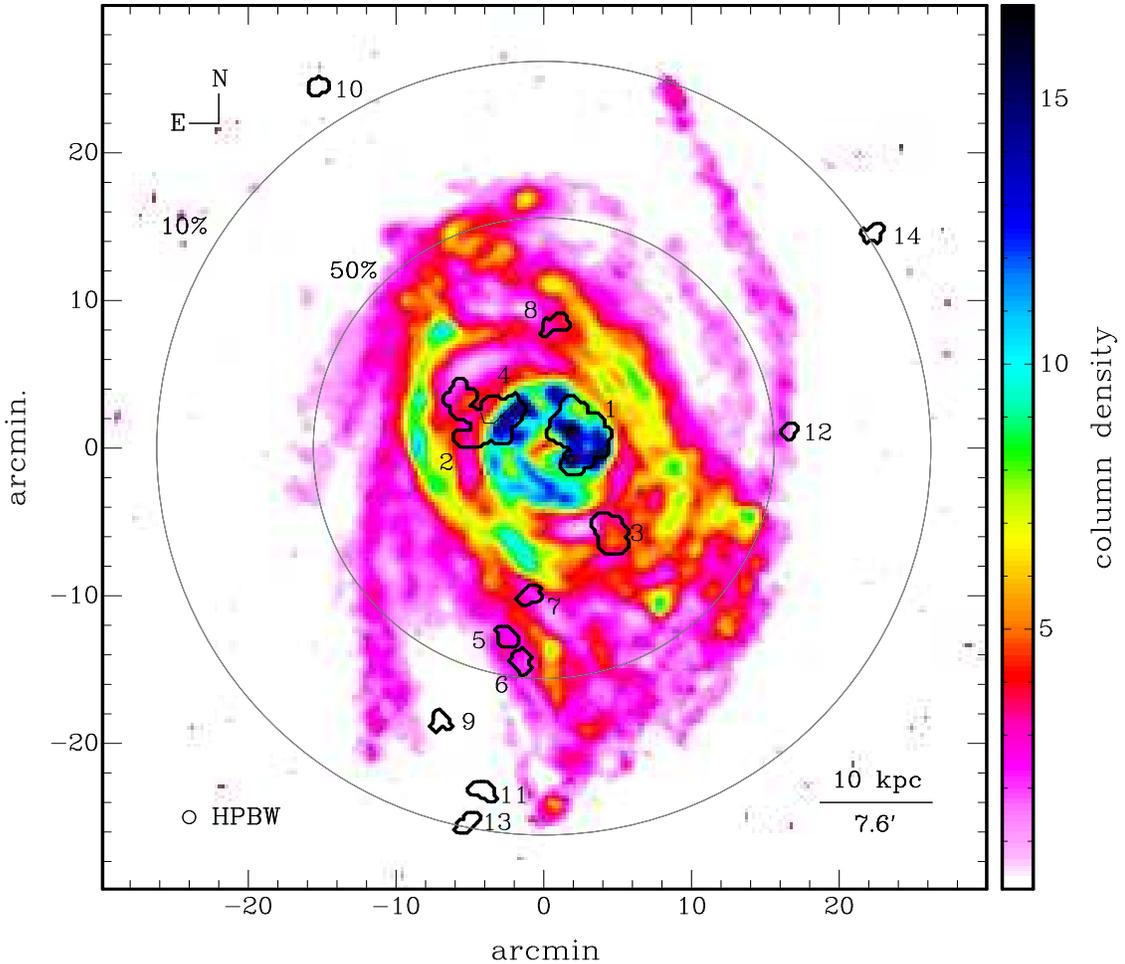}
\caption{The \hi\ column density map for M 83, showing the locations of the
detected AVCs.  
The shapes indicate the projected spatial extent of each AVC
out to 1-$\sigma$ in \hi\ column density (5\eex{18} \cm).
The large gray circles indicate the primary beam
sensitivity.  The column density is in units of \eez{20} \cm.}
\label{fig:m83mom0avcs}
\end{figure*}

\begin{figure*}
\plotone{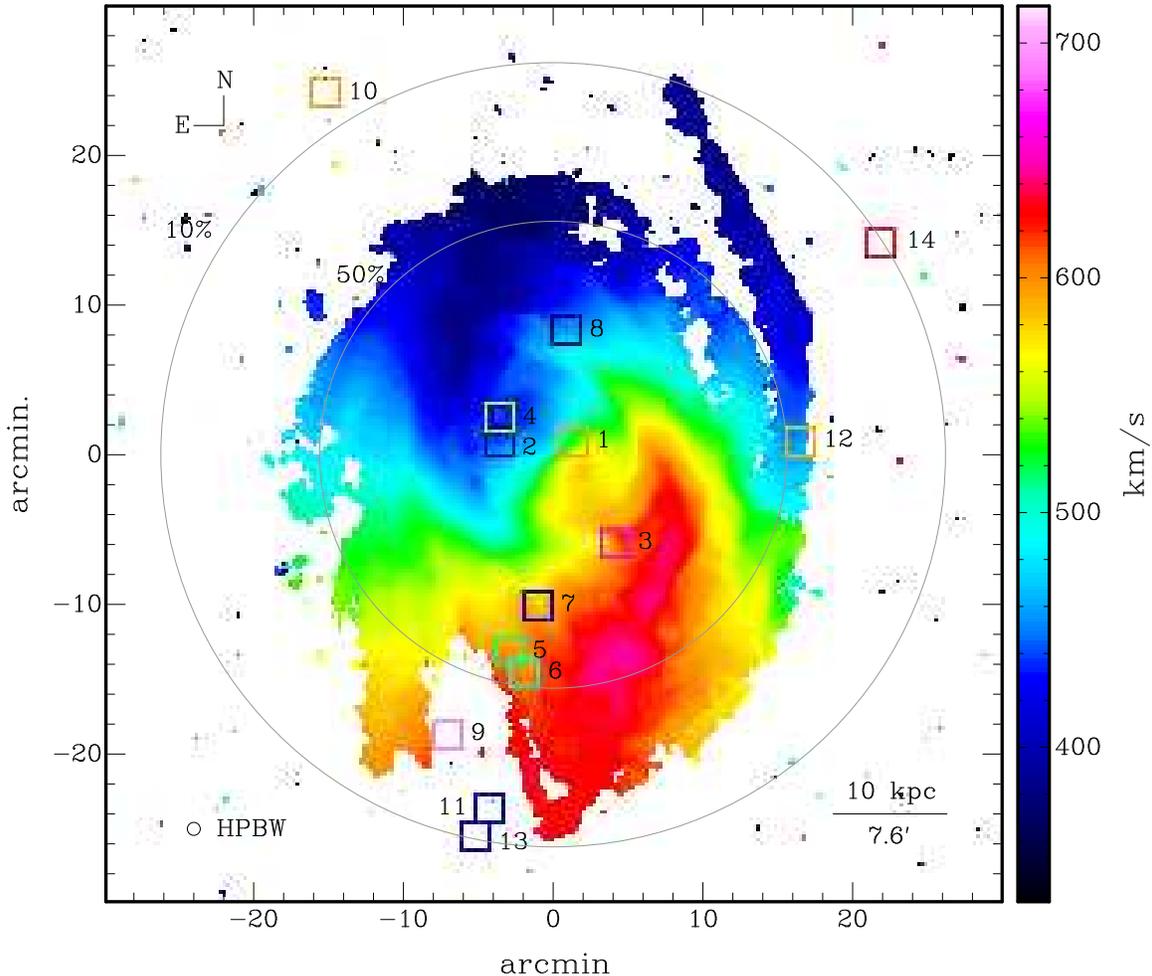}
\caption[M 83 AVCs plotted on the \hi\ velocity map]
{The \hi\ intensity-weighted mean velocity map for M 83, showing the
locations of the detected AVCs.  Each square is centered at the centroid of
the AVC, with the color indicating its heliocentric radial velocity.}
\label{fig:m83mom1avcs}
\end{figure*}

We discovered 14 discrete clouds of \hi\ emission at anomalous velocities
or positions in the M 83 data cube.  These anomalous-velocity clumps
(AVCs) are distinct from the bulk of the \hi\ disk emission in that their
contours of emission detach from the disk at levels approaching the map
noise.  The AVC properties are given in Tables \ref{tab:m83avcs1} and
\ref{tab:m83avcs2}.  The locations of the AVCs are shown superposed on the
\hi\ column density map in Figure \ref{fig:m83mom0avcs} and superposed on
the velocity map in Figure \ref{fig:m83mom1avcs}.  

\begin{deluxetable*}{ccccccccc}
\tabletypesize{\scriptsize}
\tablewidth{0pt}
\tablecaption{Discrete AVCs: General Properties
     \label{tab:m83avcs1}}
\tablehead{
\colhead{AVC\tablenotemark{a}} &
\colhead{RA} &
\colhead{Dec} &
\colhead{$v_{\rm AVC}$\tablenotemark{b}} &
\colhead{$\Omega$\tablenotemark{c}} &
\multicolumn{2}{c}{$l$\tablenotemark{d}} &
\multicolumn{2}{c}{$r_{proj}$\tablenotemark{e}} \\
\colhead{} &
\colhead{(J2000)} & 
\colhead{(J2000)} &
\colhead{(km/s)} &
\colhead{(sq.~arcmin)} &
\colhead{(\arcmin)} &
\colhead{(kpc)} &
\colhead{(\arcmin)} &
\colhead{(kpc)} 
}
\startdata
1  & 13 36 49.2 & -29 51 13 & 595$\pm\phn2$ & 2.63 & 5.3 & 6.9 & \phn1.7 & \phn2.2 \\
2  & 13 37 19.3 & -29 51 13 & 398$\pm\phn2$ & 2.75 & 5.5 & 7.2 & \phn3.7 & \phn4.8 \\
3  & 13 36 42.7 & -29 57 31 & 667$\pm\phn2$ & 0.77 & 2.8 & 3.6 & \phn7.1 & \phn9.3 \\
4  & 13 37 18.3 & -29 49 35 & 547$\pm\phn2$ & 0.58 & 2.5 & 3.3 & \phn4.4 & \phn5.7 \\
5  & 13 37 12.9 & -30 04 59 & 512$\pm\phn2$ & 0.39 & 1.7 & 2.2 &    13.4 & 17.4 \\
6  & 13 37 08.6 & -30 06 23 & 498$\pm\phn3$ & 0.39 & 1.8 & 2.3 &    14.6 & 19.0 \\
7  & 13 37 06.4 & -30 02 11 & 411$\pm\phn2$ & 0.39 & 1.5 & 2.0 &    10.0 & 13.0 \\
8  & 13 36 56.8 & -29 43 31 & 361$\pm\phn2$ & 0.39 & 1.9 & 2.5 & \phn8.4 & 11.0 \\
\hline                        
9  & 13 37 33.4 & -30 10 35 & 690$\pm\phn8$ & 0.39 & 1.3 & 1.7 &    19.9 & 25.8 \\
10 & 13 38 11.8 & -29 27 38 & 593$\pm\phn5$ & 0.39 & 1.2 & 1.6 &    28.6 & 37.2 \\
11 & 13 37 21.6 & -30 15 15 & 407$\pm12$ & 0.39 & 1.6 & 2.1 &    23.9 & 31.1 \\
12 & 13 35 45.7 & -29 50 58 & 591$\pm\phn5$ & 0.39 & 1.0 & 1.3 &    16.5 & 21.5 \\
13 & 13 37 25.9 & -30 17 35 & 405$\pm\phn5$ & 0.39 & 1.4 & 1.8 &    25.9 & 33.7 \\
14 & 13 35 19.0 & -29 37 39 & 633$\pm32$ & 0.39 & 1.2 & 1.6 &    26.3 & 34.2 \\
\enddata
\tablenotetext{a}{AVCs 1--8 are considered real detections, AVCs 9-14 are
likely spurious.}
\tablenotetext{b}{Heliocentric radial velocity of the peak of the AVC
emission.}
\tablenotetext{c}{Solid angle subtended by the AVC after collapsing its
HI emission along the velocity axis.  A value of 0.39 sq.~arcmin indicates
the feature is unresolved.}
\tablenotetext{d}{Greatest linear extent of the AVC, uncorrected for
broadening by the synthesized beamwidth.}
\tablenotetext{e}{Projected distance from the galactic center.}
\end{deluxetable*}

\begin{deluxetable*}{cccccccc}
\tabletypesize{\scriptsize}
\tablewidth{0pt}
\tablecaption{Discrete AVCs: Dynamical Properties
     \label{tab:m83avcs2}}
\tablehead{
\colhead{AVC\tablenotemark{a}} &
\colhead{$F_{\rm meas}$\tablenotemark{b}} &
\colhead{$F_{\rm corr}$\tablenotemark{c}} &
\colhead{$M_{\rm HI}$\tablenotemark{d}} &
\colhead{$N_{\rm HI}$\tablenotemark{e}} &
\colhead{$\vdev$\tablenotemark{f}} &
\colhead{$\Delta v$\tablenotemark{g}} &
\colhead{$KE$\tablenotemark{h}} \\
\colhead{} &
\colhead{(Jy km/s)} &
\colhead{(Jy km/s)} &
\colhead{(\eez{6} \msun)} &
\colhead{(\eez{20} \cm)} &
\colhead{(km/s)} &
\colhead{(km/s)} &
\colhead{(\eez{53} erg)} 
}
\startdata
1  & 3.08 $\pm$0.07 & 3.12 $\pm$0.08 &    14.9 $\pm$0.4 & \phn2.9 &   \phn$+$51 & \phn29$\pm2$  & 3.9 $\pm$0.1 \\[3pt]
2  & 1.42 $\pm$0.06 & 1.48 $\pm$0.06 & \phn7.1 $\pm$0.3 & \phn1.4 &   \phn$-$41 & \phn28$\pm2$  & 1.2 $\pm$0.0 \\[3pt]
3  & 0.99 $\pm$0.05 & 1.12 $\pm$0.05 & \phn5.4 $\pm$0.3 & \phn2.6 &   \phn$+$74 & \phn56$\pm4$  & 2.9 $\pm$0.1 \\[3pt]
4  & 0.23 $\pm$0.03 & 0.24 $\pm$0.03 & \phn1.2 $\pm$0.1 & \phn0.6 &      $+$113 & \phn32$\pm2$  & 1.5 $\pm$0.2 \\[3pt]
5  & 0.11 $\pm$0.02 & 0.18 $\pm$0.04 & \phn0.8 $\pm$0.2 & \phn1.0 &   \phn$-$77 & \phn15$\pm2$  & 0.5 $\pm$0.1 \\[3pt]
6  & 0.11 $\pm$0.02 & 0.21 $\pm$0.04 & \phn1.0 $\pm$0.2 & \phn1.1 &      $-$103 & \phn15$\pm2$  & 1.0 $\pm$0.2 \\[3pt]
7  & 0.13 $\pm$0.02 & 0.17 $\pm$0.03 & \phn0.8 $\pm$0.2 & \phn1.2 &      $-$166 & \phn44$\pm3$  & 2.3 $\pm$0.4 \\[3pt]
8  & 0.11 $\pm$0.02 & 0.13 $\pm$0.03 & \phn0.6 $\pm$0.1 & \phn0.7 &   \phn$-$90 & \phn24$\pm2$  & 0.5 $\pm$0.1 \\[3pt]
\hline
9  & 0.08 $\pm$0.02 & 0.24 $\pm$0.06 & \phn1.2 $\pm$0.3 & \phn2.3 & \phn\nodata & \phn30$\pm2$  &      \nodata \\[3pt]
10 & 0.14 $\pm$0.03 & 2.77 $\pm$0.52 &    13.2 $\pm$2.5 &    29.1 & \phn\nodata & \phn24$\pm2$  &      \nodata \\[3pt]
11 & 0.19 $\pm$0.03 & 1.09 $\pm$0.17 & \phn5.2 $\pm$0.8 & \phn9.2 & \phn\nodata & \phn82$\pm5$  &      \nodata \\[3pt]
12 & 0.06 $\pm$0.02 & 0.14 $\pm$0.04 & \phn0.7 $\pm$0.2 & \phn1.3 &      $+$135 & \phn29$\pm2$  & 1.2 $\pm$0.3 \\[3pt]
13 & 0.16 $\pm$0.03 & 1.51 $\pm$0.27 & \phn7.2 $\pm$1.3 &    13.5 & \phn\nodata & \phn20$\pm2$  &      \nodata \\[3pt]
14 & 0.15 $\pm$0.03 & 1.68 $\pm$0.31 & \phn8.0 $\pm$1.5 &    17.4 & \phn\nodata &    164$\pm12$ &      \nodata \\[3pt]
\enddata
\tablenotetext{a}{AVCs 1--8 are considered real detections, AVCs 9-14 are
likely spurious.}
\tablenotetext{b}{Total flux in the AVC, uncorrected for the primary beam
attenuation.  Errors are 1-$\sigma$.}
\tablenotetext{c}{Total flux in the AVC, corrected for the primary beam
attenuation.}
\tablenotetext{d}{Mass of HI contained in the AVC.}
\tablenotetext{e}{Peak column density of HI, integrated across the
velocity width of the emission.}
\tablenotetext{f}{Deviation velocity of the AVC, defined as the
difference between the radial velocity of the peak AVC emission and the
velocity fitted to the HI disk (if detectable) at that location, as
described in Section \ref{subsect:m83:hiavcs:clumpfind}.}
\tablenotetext{g}{Full width at half maximum of a Gaussian fit to the
velocity profile.}
\tablenotetext{h}{Kinetic energy of the AVC found from the deviation
velocity and mass.  This is a lower limit to the total kinetic energy as
only radial velocities are used.}
\end{deluxetable*}

\begin{figure*}
\plotone{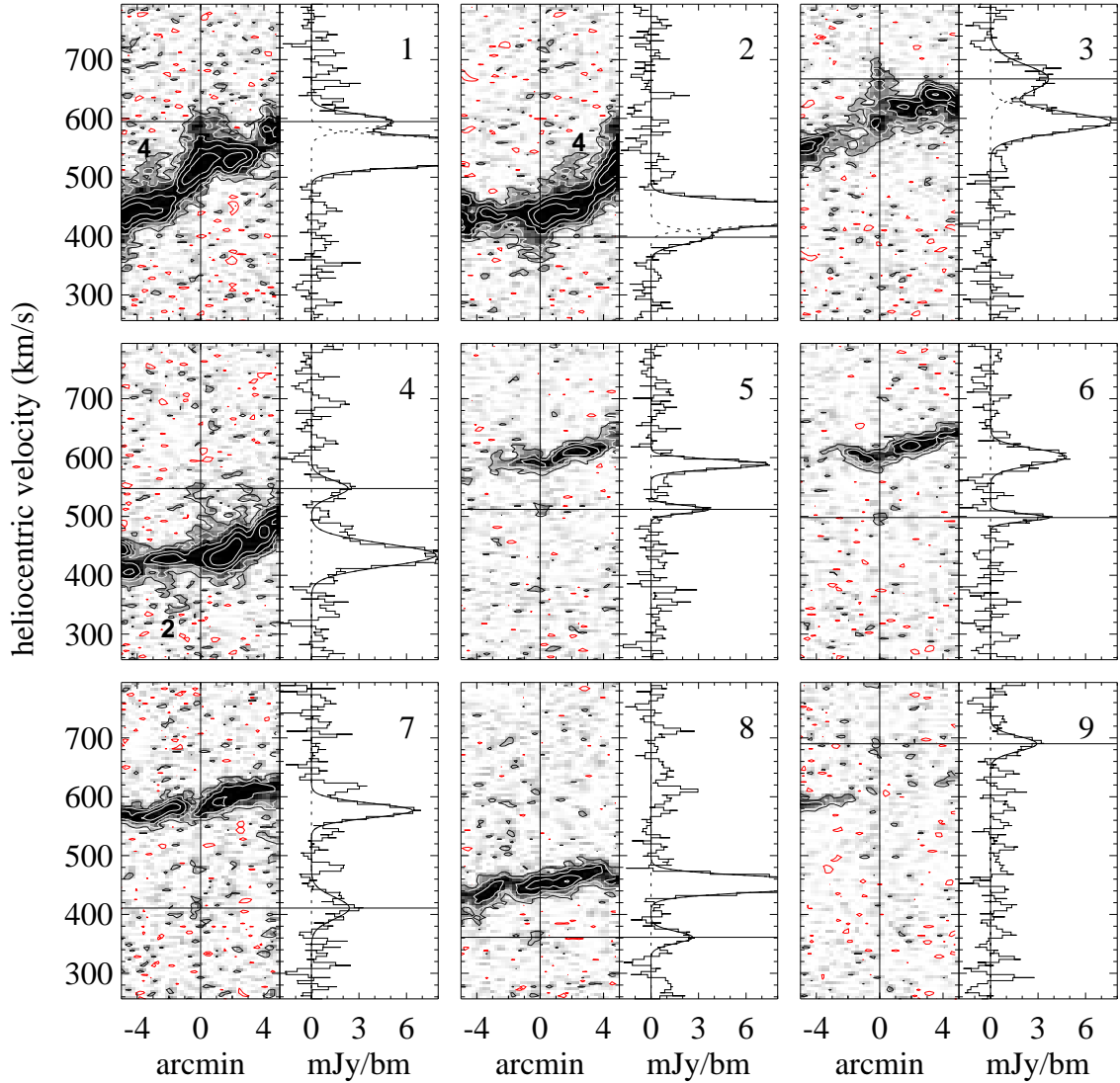}
\caption{Position-velocity slices taken along the E-W direction and
velocity profiles are shown for each AVC, which is centered at the
intersection of the solid lines in the \pv\ plot.  Grayscale ranges from 0
to 6 mJy/beam.  Black and white contours begin at 2-$\sigma$ 
($\sigma = 0.94$ mJy/beam) and increase by factors of two.  Red contours
begin at $-2$-$\sigma$ and decrease by factors of two.  The profile plot,
taken at the centroid of the AVC total \hi\ emission, shows the Gaussian
velocity fit to the AVC and \hi\ disk (if present).  AVCs which show up in
multiple slices are identified by small numerals.}
\label{fig:m83avcs}
\end{figure*}

\begin{figure*}
\plotone{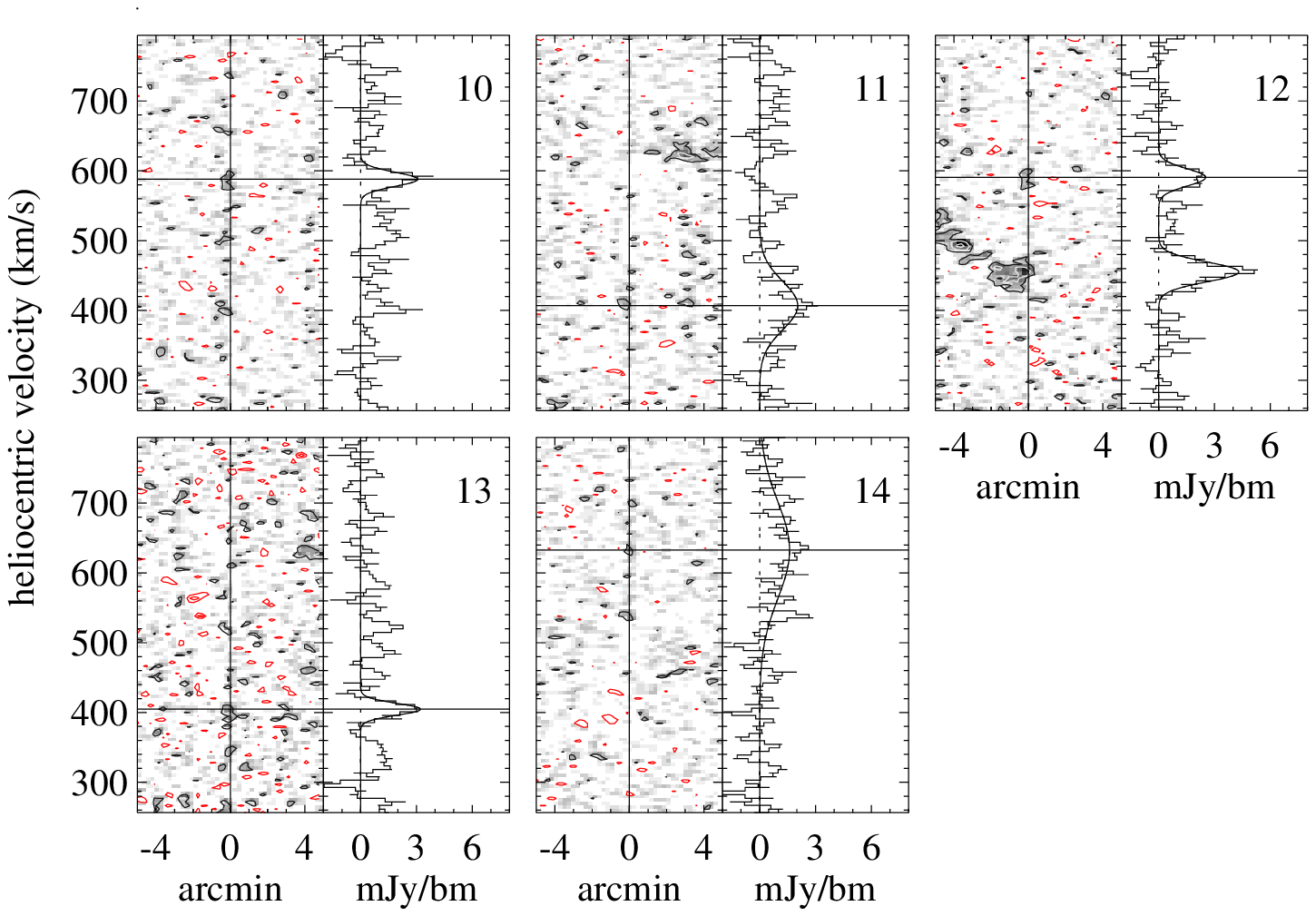}
\\
Figure~\ref{fig:m83avcs} (continued)
\end{figure*}

\begin{figure}
%\centering{\includegraphics[width=5in]{f13.eps}}
\plotone{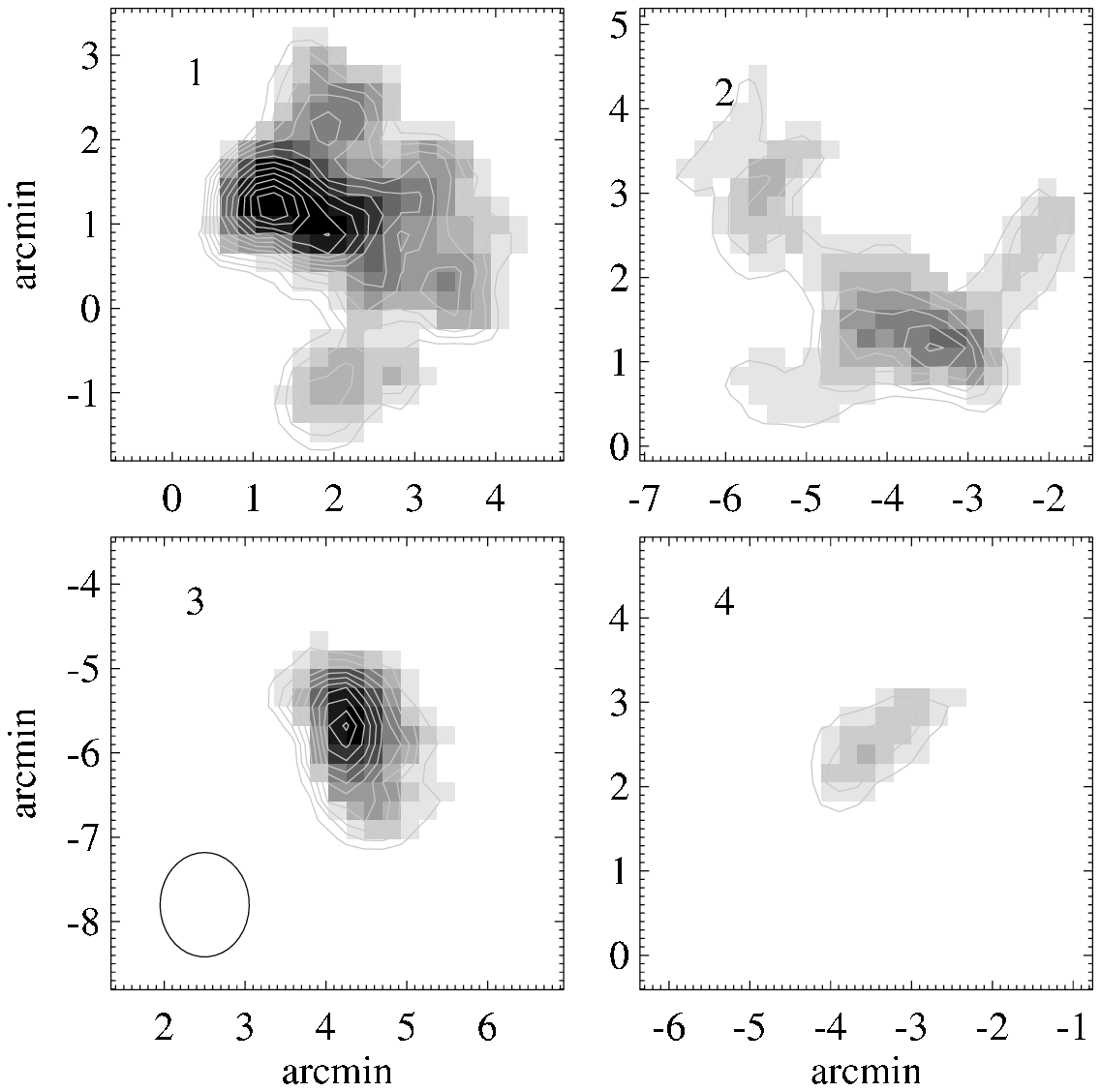}
\caption{Integrated \hi\ intensity maps for the four spatially-extended
AVCs.  AVCs 1 and 2 have bright cores and low-level extended features,
while AVCs 3 and 4 are more compact and only mildly resolved.  Grayscale
ranges from 0--0.25 Jy beam$^{-1}$ \kps, and the contours are spaced at
0.025 Jy beam$^{-1}$ \kps.  The ellipse in the AVC 3 plot shows the HPBW of
the synthesized beam.  Coordinates are with respect to the center of M 83,
with north up and east to the left.}
\label{fig:m83avcmoms}
\end{figure}

The AVCs are found across the \hi\ disk of M 83, but they do exhibit a
small degree of clustering, and several are located in regions where the
kinematics of the disk are most complicated.  Three clumps (1, 2 and 4) are
located within the kinematically ``normal'' inner disk.  One clump (3) is
found at the inner disk/warp interface, two additional sources (7 and
8) are projected just beyond the ring, and four clumps (5, 6, 9, and 12)
are found projected near outer \hi\ arms.  The remaining four AVCs (10, 11,
13, 14) are located well away from the center of the galaxy, at projected
radii where the primary beam sensitivity falls to 10\% or lower.

Gaussian fits and \pv\ plots were produced as described in the previous
section, and these are shown for each AVC in Figure \ref{fig:m83avcs}.  The
integrated \hi\ intensity maps for the spatially-extended AVCs are shown in
Figure \ref{fig:m83avcmoms}.  To eliminate the confusion of
spatially-varying noise, the \pv\ slices and intensity maps have not been
corrected for the primary beam attenuation.  The fluxes, masses, and column
densities reported in Table \ref{tab:m83avcs2} have been corrected for this
effect.

A number of the AVCs have fluxes or velocity structure in the \pv\ plots
that suggest they are spurious.  From a ``by-eye'' analysis, we estimate
that eight of these features (AVCs 1--8) are real detections and the
remainder are not.  Below we describe specific properties of individual
AVCs, grouping them according to similar distribution and attributes.  
For completeness, we include features thought to be spurious, although
these are excluded in our discussion in Section \ref{sect:m83:disc}.

\noindent
{\bf AVC 1}

\noindent
This source is the brightest and most massive AVC detected, with 
$M = 1.5\eex{7}$ \msun.  Centered 2 kpc from the dynamical center of M 83,
AVC 1 is also the only discrete clump projected over the inner \hi\ disk.
The intensity map and \pv\ plots show that it is spatially resolved and
extended in velocity, with a deviation velocity $\vdev = 51$ \kps\ and line
width of 29 \kps\ FWHM.  The measured line width is likely only accurate
near the emission peak, as the \pv\ slices display a connection in velocity
to the underlying disk emission.  The projected intensity map supports this
idea, with a secondary maximum projecting to the south, where the cold disk
approaches the velocity of the cloud.  The characteristics of this AVC are
consistent with material flowing into or out of the disk, although the lack
of spatial information along the line of sight complicates conclusions
about the source's $z$ height.  

The moment and dynamical maps described in Section
\ref{subsect:m83:hiavext:beard} clearly show emission from this feature
(see Figures \ref{fig:m83beard1} and \ref{fig:m83beard2}).  The peak of the
emission can be seen 1.5\arcmin\ west and 1.5\arcmin\ north of center in
Figure \ref{fig:m83beard1}a.  

\noindent
{\bf AVCs 2, 4}

\noindent
These two clumps are close together in projection at a radius of about 5
kpc (near the edge of the inner disk), yet they differ greatly in their
other characteristics.  AVC 2 is bright, massive ($M = 7\eex{6}$ \msun) and
extended ($\Omega = 2.8$ sq.~arcmin, $l = 7$ kpc), while AVC 4 is less
massive ($M = 1\eex{6}$ \msun) and barely resolved (largest linear extent
$l = 3.3$ kpc).  In addition, the deviation velocities of these clumps are
quite different, with AVC 2 at high relative velocity ($\vdev = -41$ \kps)
and AVC 4 at low relative velocity ($\vdev = 113$ \kps).  The line widths
are similar, 
but the emission of AVC 2 appears to blend with the disk velocity.
Emission from AVC 4 also blends kinematically with the disk, but the peak
is more clearly separated, as can be seen in the velocity profiles in
Figure \ref{fig:m83avcs}.
The heliocentric velocity
of AVC 4 (522 \kps) places it in the ``forbidden'' region of
counter-rotating material in the \pv\ plane.  The spatial coincidence of
these clumps hints at a correlation between them, possibly in the form of
an expanding bubble \citep[e.g.,][]{Kamphuisetal91}, but it is difficult
to reconcile their differing morphological and kinematic structure.  As
with AVC 1, these two clumps appear in the AV gas maps shown in Figures
\ref{fig:m83beard1} and \ref{fig:m83beard2}.

\noindent
{\bf AVC 3}

\noindent
At a projected distance of 9 kpc from the galactic center, this AVC appears
in a region where the \hi\ ring is disturbed and has a relatively low
column density of $\nh = 3\eex{20}$ \cm.  The clump is extended in space,
covering two synthesized beams in area and spanning 3.6 kpc.  It is also
extended in velocity, with a line width of 56 \kps\ FWHM, and its emission
appears to merge with the disk \hi\ at lower velocities.  The source is
massive ($M = 5.4\eex{6}$ \msun) and has a large kinetic energy of
3\eex{53} erg.

Although the clump is projected on a region of changing inclination between
the inner disk and outer arms, it does not appear to result from
beam-smearing of a warp discontinuity.  First, it is a compact structure,
whereas a warp feature would be extended like the \hi\ ring or outer arms.
Second, it is much broader in velocity than the kinematically cold \hi\
elsewhere in the galaxy.  Finally, the kinematics of the surrounding disk
material are regular in comparison to that seen at warp interfaces in other
regions of the galaxy.  This cloud is the only such feature seen along the
\hi\ ring. 

This AVC appears at a location of low disk \nh. The \pv\ plots and Figure
\ref{fig:m83mom0avcs} show a lack of emission in the disk a few arcmin to
the east of this location, although \hi\ disk emission is present at the
projected location of AVC 3.

\noindent
{\bf AVCs 5, 6}

\noindent
These two AVCs have similar mass, size, line width, and deviation velocity.
At a projected radius of about 14\arcmin\ (18 kpc) to the south of the
galactic center, AVCs 5 and 6 are separated by 2\arcmin\ (2.6 kpc) in
projected space and 20 \kps\ in velocity.  AVC 6 is the more massive and
kinematically anomalous clump, with a mass of 1\eex{6} \msun\ and deviation
velocity of $-103$ \kps. AVC 5 has a mass of 8\eex{5} \msun\ and deviation
velocity of $-77$ \kps.  The line width of both clumps is 15 \kps,
producing a velocity dispersion of 6 \kps.  Unlike the previously discussed
AVCs, the emission from these \hi\ clumps does not merge with that of the
disk in position or velocity.

In addition to similar kinematics, these two clouds have similar sizes and
are spatially unresolved by the VLA synthesized beam.  The FWHM of a
Gaussian fit across either feature is 45\arcsec; when deconvolved from the
$\sim 35\arcsec$ synthesized beam, this implies a maximum source size of
30\arcsec\ ($\sim 0.7$ kpc).  Their similar characteristics and location
hint at a possible relation, and they might be bright spots in a single
\hi\ feature, although this speculation is made less likely by the 20 \kps\
difference in deviation velocity and the small radial velocity dispersions.

\noindent
{\bf AVCs 7, 8}

\noindent
These emission sources are similar in that they appear between the \hi\
ring and an outer arm, although they are projected on opposite sides of
center.  Both have highly negative deviation velocities (\vdev\ = $-166$
and $-90$ \kps), are spatially unresolved, and have similar fluxes and
masses ($M = 8$ and 6\eex{5} \msun).  Clump 7 has a large fraction of the
kinetic energy of the AVC ensemble, due primarily to its large deviation
velocity.  From the integrated \hi\ maps, which have the same colormap
scale, one can see that AVC 7 is more sharply peaked than AVC 8.
Consideration of the \pv\ plots and velocity profiles shows that the line
of sight to AVC 7 contains emission at a variety of other velocities (e.g.,
$-100$, $-40$, +40 \kps).  The clumps producing this emission fall below
our threshold for clump identification, and it is unclear whether they
result from real velocity structure in the AVC or from a systematic problem
such as poor continuum subtraction or bandpass calibration.

\noindent
{\bf AVCs 9, 12}

\noindent
Two clumps lie superposed on or near \hi\ arms outside the half power width
of the primary beam.  Both are spatially unresolved and have similar line
widths of 30 \kps\ FWHM.  At these radii, the primary beam attenuation
correction factor is about 3, so although both have small measured fluxes,
the corrected masses are significant (1.1 and 0.6\eex{6} \msun).  Both AVCs
have high deviation velocities, with $\vdev = +135$ \kps\ for clump 12, and
a velocity difference of $+100$ \kps\ for clump 9 measured from the \hi\
arm projected nearby.  These detections are the least statistically
significant of the set (4-$\sigma$ and 3-$\sigma$ for AVC 9 and 12,
respectively), and they are hereafter treated as spurious.

\noindent
{\bf AVCs 10, 11, 13, 14}

\noindent
All four of these clumps are in low-sensitivity regions, where the
attenuation correction factor approaches 10.  AVCs 11 and 13 are projected
close together in space (2\arcmin\ apart) and velocity (2 \kps\ apart).  All
four have strongly peaked emission maps, and their lines of sight contain
emission at velocities throughout the data cube.  In addition, AVCs 11 and
14 have very large line widths compared to the other AVCs.  It is likely
that these AVCs are not real, but are the result of systematic effects such
as calibration errors.  If they are real, the beam correction produces
large masses and kinetic energies for these clumps.

\subsection{Optical Counterparts to the AVCs}

\begin{figure*}
\plotone{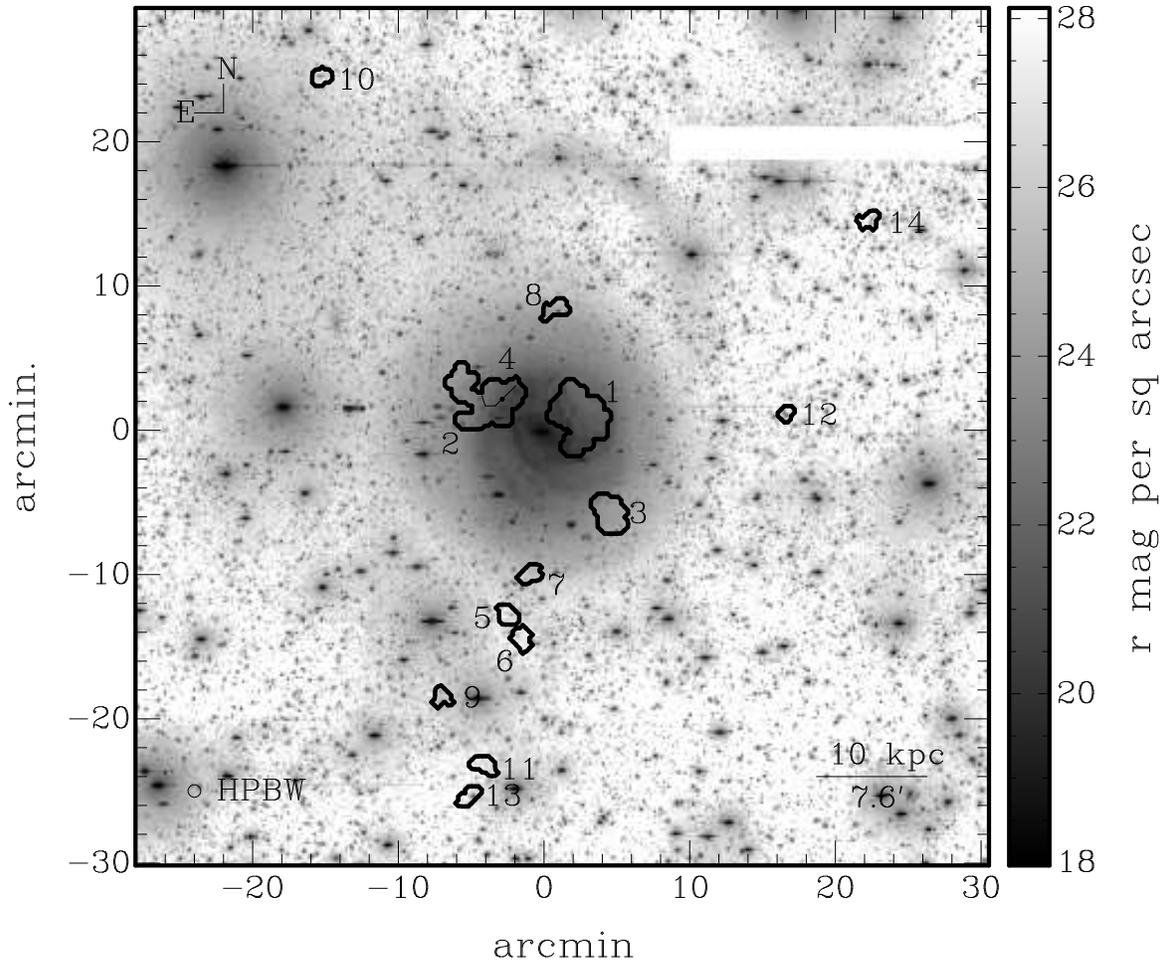}
\caption{The optical M 83 image overlaid with the locations of the \hi\
AVCs.}
\label{fig:m83optavcs}
\end{figure*}

Searching for optical counterparts to the \hi\ AVCs was carried out by
hand.  Each AVC was overlaid on the unmasked optical image, which was
inspected for optical emission (see Figure \ref{fig:m83optavcs}).  To
search for small-scale emission structure superposed on the extended galaxy
profile, the masked optical image was median filtered with a $5\times5$
pixel box.  The location of each AVC was fit with a plane, excluding the
four AVCs within the inner disk, which is difficult to fit with a simple
model.  The fit residuals were inspected for signal above the noise, which
was not found in any case.   The noise in the residuals was used to
estimate surface brightness upper limits for optical counterparts.  The
results are shown in Table \ref{tab:m83optavcs}.

\begin{deluxetable}{cccc}
\tabletypesize{\scriptsize}
\tablewidth{0pt}
\tablecaption{Optical Counterparts to HI AVCs
     \label{tab:m83optavcs}}
\tablehead{
\colhead{AVC\tablenotemark{a}} &
\colhead{$\mu_{\rsloan,0}$\tablenotemark{b}} &
\colhead{$\mu_{\rsloan,{\rm lim}}$\tablenotemark{c}} &
\colhead{Notes\tablenotemark{d}}
}
\startdata
1   & 20.4  & \nodata  & inner disk, spiral arm \\
2   & 20.7  & \nodata  & inner disk, spiral arm \\
3   & 24.2  & \nodata  & inner disk \\
4   & 22.2  & \nodata  & inner disk \\
5   & 27.8  & 27.2  & stars \\
6   & 27.9  & 27.4  & stars \\
7   & 25.4  & 27.1  & stars \\
8   & 24.4  & 26.4  & outer disk, stars \\
\hline
9   & 26.5  & 27.0  & outer disk, stars \\
10  & 26.8  & 27.3  & stars, galaxy 1.5\arcmin\ SE \\
11  & 27.2  & 26.8  & stars \\
12  & 27.5  & 26.8  & bright coincident stars \\
13  & 28.1  & 27.4  & stars \\
14  & 27.7  & 26.9  & stars, galaxy \\
\enddata
\tablenotetext{a}{AVCs 1--8 are considered real detections, AVCs 9-14 are
likely spurious.}
\tablenotetext{b}{The median surface brightness, in \rsloan\ magnitudes per
square arcsec, at the location of the AVC.  For extended AVCs, the median
was taken within the HPBW centered on the peak HI emission.}
\tablenotetext{c}{The 3-$\sigma$ surface brightness limit on any optical
features coincident with the AVC.  This was determined by fitting and
subtracting a plane from the optical image at that point.  See that text
for details.}
\tablenotetext{d}{The notes identify optical features coincident with the
AVC, including foreground stars, extended emission from M 83 (designated by
location in the disk), and background galaxies.}
\end{deluxetable}

Below we describe possible counterparts and limits to the optical surface
brightness for the individual AVCs.  We discuss these limits in the context
of other deep optical studies in Section \ref{sect:m83:disc:accret}.

\noindent
{\bf AVCs 1, 2, 3, 4, 7, 8}

\noindent
These AVCS are all projected over the bright optical disk.  The bright cores
of AVCs 1 and 2 are projected on spiral arms, appearing to trace the shape
of the arms, although this is possibly due to the method used to detect and
delineate the \hi\ clumps.  The optical light coincident with AVC 1 peaks
at 20.4 \rsloan\ mag per square arcsec, excluding the bright star clusters
in the spiral arms, and falls off to 22.1 \rsloan\ mag per square arcsec in
the outer part of the clump.  Likewise, the core of AVC 2 is projected on a
bright spiral arm, with $\mu_{\rsloan}$ = 20.7 \rsloan\ mag per
square arcsec.  The optical light falls to 24.5 \rsloan\ mag per square
arcsec in the outer regions of this extended \hi\ clump.  While these AVCs
align with spiral arms that contain star clusters, the spatial resolution
of the \hi\ data is insufficient to draw direct comparisons between AVC and
star cluster morphology.

The other two AVCs within the inner disk, AVCs 3 and 4, have no obvious
small-scale optical counterparts aside from stars.  The inner edge of AVC 4
coincides with the outer edge of a spiral arm, although this is again
likely due to the searching and masking technique we have employed.  AVC 3
is projected on a disruption in the \hi\ disk, but no evidence of such a
feature is seen in the optical data.

Clumps 7 and 8 are projected further out, over the south and north edge of
the extended stellar envelope.  The underlying surface brightnesses in
these regions is 25.4 and 24.4 \rsloan\ mag per square arcsec,
respectively.  There are no small-scale optical features, aside from stars,
down to the detection limit of about 27 \rsloan\ mag per square arcsec.

\noindent
{\bf AVCs 5, 6, 9, 11, 13}

\noindent
The majority of the large-radius AVCs contain only stars within the lowest
\hi\ contour.  Clumps 5 and 6 are isolated from bright stars, but AVCs 9,
11 and 13 lie in the wings of very bright stars.  As a result, much of the
light in these regions is masked out to determine the surface brightness, and
the amount of flux below a limiting magnitude of about 26 \rsloan\ mag per
square arcsec is poorly constrained.  There is no evidence for emission
down to this level, however.

\noindent
{\bf AVCs 10, 12, 14}

\noindent
These three AVCs are positioned on or near optical features that could be
related.  AVC 10 is projected 1.5\arcmin\ northwest of an object with
a galactic light profile.  The contours of AVC 14 contain a similar object.
Both objects are smaller than the extent of the AVCs, although
the \hi\ clumps are unresolved.  Given their apparent size, it is likely
that these optical sources are background galaxies, although it is
impossible to verify without spectroscopic data.

Clump 12 is centered on the location of two bright stars, making the
limiting magnitude at this location difficult to determine.  There is no
signal at this position in the 1.4 GHz continuum map, therefore it is
unlikely that the detected \hi\ results from poor continuum subtraction.
It is more likely that this is a chance superposition between bright stars
and a detected \hi\ source (real or spurious).

\subsection{AVC Detection Simulations} 
\label{subsect:m83:hiavcs:sim}

To analyze our results quantitatively, it is necessary to determine
significance of the detections in terms of the completeness and false
detection rate.  Interferometry maps differ from images made with
non-interferometric devices in that neighboring spatial pixels are not
independent.  The transformation of limited samples from the $uv$ plane to
the image plane results in the signal from every location being spatially
smeared.  Resolution elements therefore overlap each other, and more
importantly, noise features are extended and often similar in size to the
objects being searched for.  Prior to primary beam correction, the noise is
nearly constant across the field.  It is upon this background of
spatially-dependent fluctuations that one searches for discrete emission
features.

\begin{figure}
\plotone{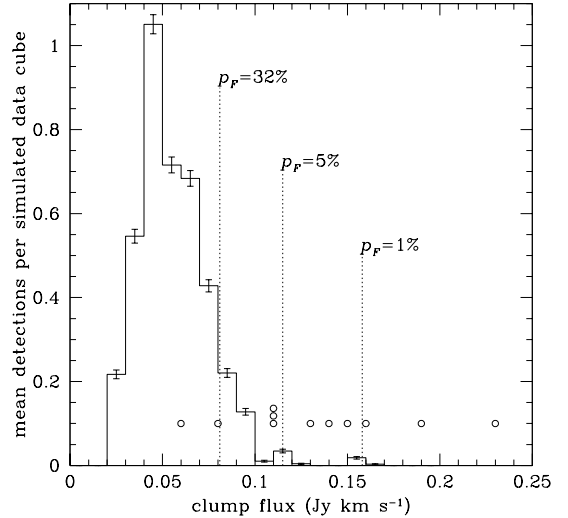}
\caption {Flux distribution of false detections in the simulated pure noise
data cubes.  The binned values have been scaled to equal the mean number of
false detections per masked M 83 data cube per 0.01 Jy \kps\ flux bin, so
that the integral of this distribution equals the mean expected number of
false detections in our data cube, a value of 4.1.  The dotted lines note
the likelihood $p_F$ of obtaining at least one false detection above the
given flux value in a single data cube (see text for further explanation).
Open circles indicate the fluxes of detected clumps in the M 83 data, with
the three brightest clumps (AVCs 1--3) excluded.  For clump flux $>$ 0.1 Jy
\kps, there is an excess over purely spurious detections.}
\label{fig:m83simfalse}
\end{figure}

To investigate the sensitivity of our data, we simulated 2000 observations
of Gaussian noise in the $uv$ plane, using the same $uv$ distribution and
noise characteristics as our filtered M 83 observations.  Each dataset was
transformed to the image plane using the same technique described in
Section \ref{sect:vlaobs}, so that the resultant image cube had dimensions,
spatial coherence (i.e., synthesized beamwidth) and $\sigma$ identical to
that of the cleaned, binned M 83 data cube.  Each simulated cube was
processed with our detection software, using identical parameters as
described in Section \ref{subsect:m83:hiavcs:clumpfind}, except for a
single filtering kernel of 5 channels (25 \kps) FWHM, the average linewidth
of the AVCs.  The histogram of clump flux values, scaled to the parameter
space volume of a single masked M 83 data cube, is shown in Figure
\ref{fig:m83simfalse}.  The low-flux side of the histogram shows a rapid
fall-off due to decreasing detection efficiency.  The histogram peaks near
0.045 Jy \kps, which is about 4.5 times the 1-$\sigma$ flux for a single
beam in the smoothed cube (see Appendix \ref{app:snrch}).  This is
consistent with our detection threshold of 4-$\sigma$.

We can use the simulated distribution to calculate the likelihood of
obtaining at least one spurious detection above a given flux threshold in
the data cube.  If we assume the false detection rate is determined by
Poisson statistics, we find that the likelihood of obtaining at least one
detection above flux $F$ is given by 
$p_F = 1 - e^{-\mu_F} \mu_F^{\nu}/\nu!  = 1 - e^{-\mu_F}$ for $\nu = 0$.
The mean expected number of detections $\mu$ is given by the integral of
the empirical distribution between $F$ and $+\infty$, and it is a
straightforward matter to find $F$ for various values of $p_F$.  The flux
limits for $p_F$ = 0.32, 0.05, and 0.01 are plotted in Figure
\ref{fig:m83simfalse}.  Several of our AVCs have fluxes (uncorrected for
attenuation) within the region of non-zero false detection rate.  If we
exclude the three significant clumps with uncorrected fluxes near or above
1.0 Jy \kps\ (AVCs 1--3), we are left with 11 clumps in the 0.06--0.23 Jy
\kps\ flux range.  These are indicated in Figure \ref{fig:m83simfalse}.  

\begin{figure}
\plotone{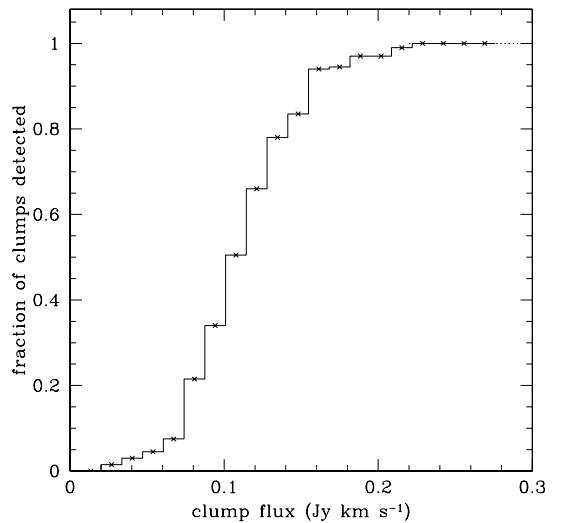}
\caption{Detection efficiency of the simulated clumps.  The efficiency is
90\% above 0.15 Jy \kps.}
\label{fig:m83simeffic}
\end{figure}

Integrating the full empirical distribution leads to an expected mean of
4.1 false detections per masked data cube.  With an algorithm flux
detection threshold of 4-$\sigma$, and for 177,500 independent samples in
our masked data cube, we expect 5.3 detections of greater than 4-$\sigma$
significance from purely statistical considerations.  That the simulations
average fewer false detections indicate limitations of our detection
algorithm.  Both false detection rates are consistent with our ``by-eye''
analysis indicating that AVCs 9--14 are spurious.  The chance of obtaining
at least 6 spurious detections given the simulation mean of 4.1 is 22\%;
given the statistical mean of 5.3, the chance is 44\%.

We note that the simulations do not account for systematic errors which may
occur from bandpass calibration or continuum subtraction and which may
introduce artifacts similar to clumps of emission.  Some of our
``detections'' show indications of this effect (e.g., the multiple velocity
components of AVCs 11 and 14).  Excluding these, the number of false
detections is still consistent with the expected rate.  Modeling of
systematic calibration errors will be included in future versions of the
simulation software.

The completeness of our sample was determined from the detection
efficiency, which in turn was found from additional simulations. Twenty
data cubes with noise characteristics similar to the data were created in
the manner described above.  To each of them were added 200 3-d Gaussian
sources with spatial FWHM = 40\arcsec\ (approximately the synthesized
beamwidth), velocity FWHM = 5 channels (25 \kps),  and peak brightness
varying between 1.0 and 5.0 mJy/beam in steps of 0.5 mJy/beam.  The
integrated fluxes of the sources varied between 0.01 and 0.27 Jy \kps.  The
positions of the sources were produced semi-randomly so that the centers
were not necessarily at integer pixels but the sources did not overlap
within 5-$\sigma$ in any direction.  The resulting cubes were processed
with our detection software in an identical fashion as the source data.

At least 90\% of the simulated sources were detected to a flux of 0.16 Jy
\kps, as is shown in Figure \ref{fig:m83simeffic}.  Below this, the
detection efficiency falls off rapidly, reaching 50\% at a flux of about
0.1 Jy \kps.  The shape of the efficiency curve may explain the lack of
low-flux detections in our sample.  Also from these simulations, we note a
systematic flux overestimate of $\sim$ 10\% for a simulated 0.2 Jy \kps\
clump and $\sim$ 30\% for 0.1 Jy \kps.  The measured AVC fluxes have not
been corrected for this effect.

%%%%%%%%%%%%%%%%%%%%%%%%%%%%%%%%%%%%%%%%%%%%%%%%%%%%%%%%%%%%%%%%%%%%%%%%
% SECTION -- Discussion
%%%%%%%%%%%%%%%%%%%%%%%%%%%%%%%%%%%%%%%%%%%%%%%%%%%%%%%%%%%%%%%%%%%%%%%%
\section{DISCUSSION}
\label{sect:m83:disc}

Different HVC production scenarios predict different characteristics for
the HVCs.  The galactic fountain predicts that material will be projected
on the disk, while tidal stripping and accretion predict HVCs across the
field of view.  We can divide the detected AVCs into two categories: those
projected on the disk (AVCs 1, 2 and 4 and the low relative velocity
extended emission), and those projected away from the disk.  Separate
treatment of each group places constraints on the importance of the various
HVC formation schemes.

\subsection{Galactic Fountain: ``The Beard'' and HVCs}
\label{sect:m83:disc:beard}

Under the galactic fountain model, hot gas is ejected from the disk by
multiple supernova explosions in star-forming regions.  This material rises
at roughly the adiabatic sound speed, moves outward due to a decrease in
the radial gravitational potential, and rotates more slowly due to
conservation of angular momentum.  After the gas cools, it loses its
buoyancy and spirals back to the disk, either as small clouds or sheets
\citep{ShapiroField76,Bregman80}.

The observational implications of this model depend in detail on the nature
of the hot corona and magnetic field, which are ill-constrained, but
several general features should be apparent in observations of external
galaxies.  First of all, discrete clouds of scale $\sim$ 1 kpc should be
observable at velocities between systemic and systemic plus the coronal
sound speed ($\lesssim 150$ \kps), on both wings of the disk velocity
profile.  Second, \hi\ emission could appear at velocities between systemic
and the disk, i.e., on the low-relative velocity wing of the disk profile.
This emission could be spatially extended (spanning a few kpc) if
small-scale condensations are suppressed
\citep{Field65,MathewsBregman78,ChevalierOegerle79,Bregman80}.  Finally,
anomalous velocity \hi\ should coincide with the spiral arms and \halpha\
disk emission, especially the discrete high-velocity clumps, which would be
infalling near their point of origin.

Our results are consistent with these predictions.  The extended anomalous
disk of \hi\ appears to be rotating more slowly than the main \hi\ disk,
which would be the case for emission from a galactic fountain.  
This material must be spatially separated from the bulk \hi, and we
conclude that it lies in a vertically extended disk.
A similar
phenomenon has been observed in a number of inclined spiral galaxies,
including NGC 891 \citep{SwatersSancisivanderHulst97,Oosterlooetal2007},
NGC 2403 \citep{Fraternalietal02a}, NGC 4559 \citep{Barbierietal2005}, 
NGC 253 \citep{Boomsmaetal2005}, 
and NGC 6946 \citep{Boomsmaetal2008}.
In particular, for the the $i = 60\arcdeg$ spiral NGC 2403,
\citet{Fraternalietal02a} argue that
this emission, termed the ``beard'' due to its appearance in \pv\ diagrams,
arises from a vertically extended \hi\ component that is rotating 20--50
\kps\ more slowly than the cold disk.  They model the anomalous disk and
characterize it as having a radial inflow at velocity $\lesssim$ 20 \kps,
implying an inflowing mass rate of $\mdot~\sim~0.3$--0.6~\msun~yr$^{-1}$
for an infall timescale of $\sim \eez{8}$ yr.  They also detect
\eez{6}--\eez{7}~\msun\ ``streams'' and ``spurs'' of \hi\ emission, ranging
in size from 5--10 kpc and deviation velocity from 50--60 \kps.  They
conclude that this material arises from a galactic fountain, with a mass
exchange rate close to the predicted value of $\sim$ 1 \msun~yr$^{-1}$.
Further evidence is provided by {\it Chandra\/} X-ray observations showing
an extended corona of $T \sim 5\eex{6}$ K gas with a radiative cooling rate
of 0.1--0.2~\msun~yr$^{-1}$ \citep{Fraternalietal02b}.  Additional modeling
by \citet{FraternaliBinney2006} supports this conclusion, although it
suggests that some amount of IGM accretion must be present to temper the
fountain and produce the low angular momentum material.

The face-on inclination of M 83 makes it difficult to model the rotation of
the beard component in the same way as \citet{Fraternalietal02a},
\citet{FraternaliBinney2006}, and others.  It is also not clear whether the
anomalous disk is radially infalling as is the case of NGC 2403, nor is it
clear how vertically extended the material is.  If the beard is the result
of a galactic fountain, it is likely that it represents material near
maximum $z$ because of its low velocity, and it probably is gas that has
recently condensed and is beginning to fall back onto the disk.  
The discrete clouds projected on the disk at higher deviation velocities
would be in freefall at lower $z$.  If the column density of a cloud
exceeds that of the disk, the cloud would punch through and disrupt the
\hi\ disk in that location.  AVC 3 is projected over a region of low disk
\nh, however it is projected near the warp and not the spiral arms, so it
may have a different origin than the other disk-coincident AVCs.

The dynamical properties of the AV gas are consistent with what would be
expected in a galactic fountain.
The beard mass of 5.6\eex{7}~\msun\ (8.4\eex{7}~\msun\ including the AVCs)
and an estimated cooling/freefall time of 5\eex{7} yr leads to a mass
exchange rate of 1~\msun~yr$^{-1}$, similar to the predictions of the
model \citep{Bregman80}.
The star formation rates of M 83 and the Milky Way are similar, so one
would expect similar mass exchange rates.  \citet{Talbot80} estimates a
star formation rate (SFR) in the Milky Way between 1.7~\msun~yr$^{-1}$ (from
converting CO measurements to H$_2$) and 2.0~\msun~yr$^{-1}$ (from \halpha).
\citet{BellKennicutt01} constrain the the SFR of M 83 to be between
1.1--2.4~\msun~yr$^{-1}$ from FUV and \halpha\ data, using the relations of
\citet{Kennicutt98}.  The mass infall rates in both galaxies are estimated
to be $\sim$ 1~\msun~yr$^{-1}$, consistent with the similar SFR.
The kinetic energy of the discrete AVCs projected
over the disk total about 7\eex{53}~erg, equaling the total output of
about 1000 supernovae.  This is a lower limit to the kinetic energy
initially imparted to the gas by the collection of supernovae, since the
velocity is measured in only one direction and the $z$ height of the AVCs
is unknown.  A supernova rate of 0.01~yr$^{-1}$ is consistent
with this if only 1\% of the energy is converted to kinetic energy.

\subsection{Galactic Disruption and Accretion: External HVCs}
\label{sect:m83:disc:accret}

It is difficult to reconcile the presence of AVCs projected outside the
stellar disk with the galactic fountain model, thus there is likely another
HVC production scenario at work.  AVCs 3, 7 and 8 are projected on the \hi\
warp/ring, and are possibly associated with that.  The remainder of the AVCs
are compact, and all except AVCs 5 and 6 are likely spurious detections.

The optical results rule out emission from a population of dwarf galaxies
similar to those in the Local Group.  The 3-$\sigma$ detection limits for
the AVCs projected outside the \hi\ disk are at least 26.8 \rsloan\
magnitudes per square arcsec.  Assuming an integrated color \vr\ $\sim$
0.4--0.6, typical of Local Group dwarfs \citep{Mateo98}, and using the SDSS
filter transformation from \citet{Smithetal02}, this corresponds to about
27.2 \V\ magnitudes per square arcsec.  This limit is faint enough to
detect the center of any of the dwarfs tabulated by \citet{Mateo98}.
Likewise, the inferred ratios of $\mhi/L_V > 2 \msun/\lsun$ are larger than
those of the majority of Local Group dwarfs, indicating a dearth of
starlight compared to these systems.

Our optical depth is similar to that of other HVC studies.
\citet{Willmanetal2002b} use SDSS data to search for stellar components in
13 Galactic HVCs, detecting none to limits of 26.7--30.1 \V\ magnitudes per
square arcsec. \citet{Simonetal2006} obtain an upper limit of 25.25--26.25
\V\ magnitudes per square arcsec for HVC Complex H, depending on the
assumed stellar population.  Fainter companions and stellar streams are
observed in the Milky Way and M 31, but they do not have evidence of \hi\
emission \citep[e.g.,][]{Zuckeretal2004,Willmanetal2005}.

If they are real, these \hi\ clouds are best explained by a combination of
tidal disruption and accretion.  The \hi\ warp, \hi\ ring, and elongated,
gas-free companion KK208 are all evidence supporting a tidal interaction at
some point in the galaxy's past.  In addition, an unpublished wide-field
\hi\ map shows evidence of an \hi\ arm to the north of the galaxy,
extending east from the tip of the longest NW arm visible in
Figure~\ref{fig:m83mom0} \citep{Parketal01}.  Our field of view is
insufficient to include this region, but it is further evidence of a tidal
interaction that can strip diffuse material and produce anomalous \hi\
emission.

\subsection{Limits on the Mass Distribution of Galactic HVCs}
\label{sect:m83:disc:mwcomp}

\begin{figure}
\plotone{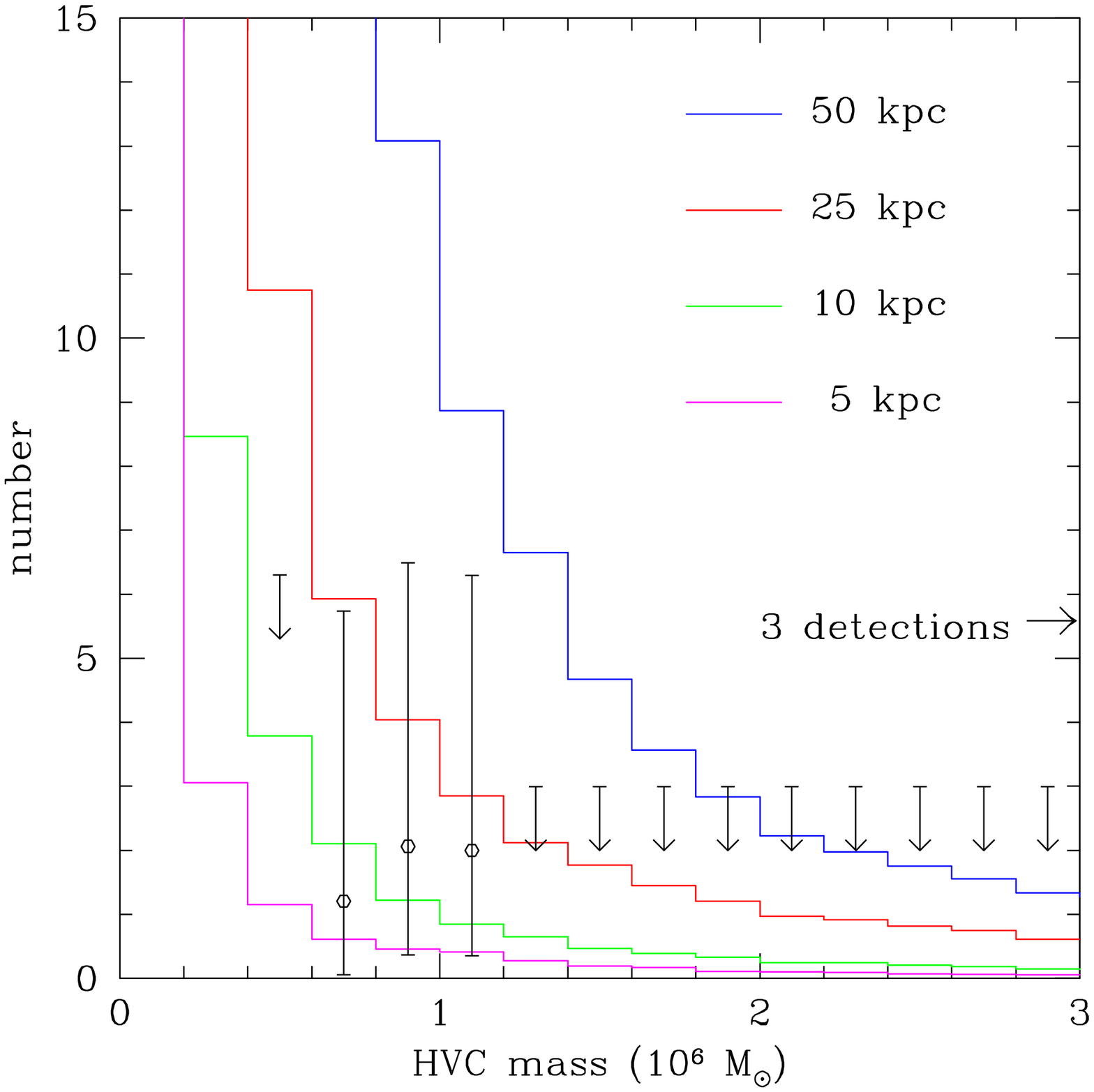}
\caption{Mass distribution of the Milky Way HVCs, assuming uniform
spherical distributions of varying mean distance and using the HVC catalog
of \citet{WvW91}.  The points show the distribution of our lowest-mass
detections in M 83, with 95\% confidence errorbars and upper limits.  
Three detections lie off the horizontal scale at $\mhi > 5\eex{6}$ \msun.
This distribution has been corrected for the false discovery rate (the
``background'') and the source detection efficiency.  If the underlying HVC
mass distributions of the two galaxies are identical, the Milky Way HVCs
must be closer than about 25 kpc or we would have detected more HVCs in M
83.}
\label{fig:m83mwmasscomp}
\end{figure}

Our results are sensitive enough to place constraints on the mass
distribution of the HVC ensemble in M 83 and, by extension, the Milky Way.
Because of the variable mass sensitivity across the field, we restrict our
analysis to AVCs detected within the HPBW of the primary beam.  We assume
that the full distribution of M 83 HVCs falls within this region, which is
a reasonable assumption if the vertical distance $z$ is less than about 50
kpc (and $z \sin{i} \lesssim 17$ kpc or 15\arcmin).  Only the eight confirmed 
detections lie within the primary HPBW, and we scale our estimates of the
false detection rate to the reduced number of independent sightlines we are
now sampling.  Five of the eight detected clumps have masses in a region
where the false detection rate is significant (greater than 0.1 detection
expected per mass bin), while the remaining AVCs are much more massive and
unlikely to be spurious.  Assuming a Poisson distribution for false
detections, and using the integrated false detection rate as the mean
``background'' expected for a mass bin, we determine the
background-corrected mass distribution with 90\% confidence limits.  This
is shown in Figure \ref{fig:m83mwmasscomp}, using the tabulated data of
\citet{Gehrels86} to calculate confidence limits for low number counts.  We
obtain upper limits above an HVC mass of $1.2\eex{6} \msun$ and below a
mass of $6\eex{5} \msun$; below $4\eex{5} \msun$, the completeness falls
below 50\%, so we have cut off the distribution at this point.

To draw conclusions about the Galactic HVC population, two simplifying
assumptions must be made.  First, we assume that the underlying HVC mass
distributions are identical for the Milky Way and M 83.  Second, we assume
the Galactic HVCs are contained in a uniform spherical distribution about
the Galactic center.  While this is likely incorrect, it is a sensible
choice given our lack of knowledge of their distances.  The collection of
Galactic HVCs across the whole sky has been cataloged by \citet{WvW91}, and
excluding the Magellanic Stream and outer Galactic arm, we randomly assign
distances drawn from the assumed distribution to the cataloged HVCs.  The
assigned distances and cataloged \hi\ fluxes produce an estimate of the
Galactic HVC mass distribution.  This procedure was repeated
500 times to fill in the high-mass regions, and the resulting distribution
was scaled to the number of HVCs in the catalog (560).  

The derived HVC mass distribution is plotted in Figure
\ref{fig:m83mwmasscomp} for four different mean Galactic HVC distances,
along with the HVC mass distribution for M 83.  While the confidence limits
are large, it is clear that an ensemble distance of less than about 25 kpc
is consistent with the M 83 distribution, while a larger average distance
results in a larger number of massive clouds.  Such clouds would have been
easily visible in our M 83 observations.  It is possible that some HVCs in
M 83 could fall outside the HPBW and therefore be missed if the distance is
much greater than 50 kpc, however even at 50 kpc we detect an order of
magnitude fewer HVCs than should be present in M 83.  In addition, the
heliocentric distances are upper limits to the more fundamental $z$ height
of the HVCs, therefore the distances of the HVCs from the plane are
consistent with being less than about 10 kpc.  Finally, if the HVCs
cataloged by \citet{WvW91} are not distinct but rather clumps within larger
complexes, then the mass distribution derived here will underestimate the
number of high-mass HVCs, and the constraints imposed by our results would
be stronger.

These constraints are consistent with previous results.  \citet{Wakker01} has
published a catalog of 18 HVCs and 8 IVCs with distances and metallicities
constrained by absorption against background stars or extragalactic
sources.  The metallicities vary from solar to a few percent of solar.  The
upper bounds to the vertical $z$ heights range from 0.1 kpc to 7 kpc, and
the corresponding mass limits range from about \eez{4} to 2\eex{6} \msun.
More recent results indicate distances of 5--15 kpc to a handful of
additional Galactic HVCs
\citep{Thometal2006,Wakkeretal2007,Wakkeretal2008}.
Similarly, \citet{Putmanetal2003} have detected \halpha\ emission in 25 HVCs
and use this to limit the distance of these objects from the ionizing
Galactic radiation field and thus constrain their height above the Galactic
plane.  The values range from $5 < z < 40$ kpc.  
Limits from external galaxies are similar.
\citet{Thilkeretal04} detected an ensemble of 20 discrete \hi\ clouds
within 50 kpc of M 31.  These sources have \hi\ masses in the range
\eez{5}--\eez{7} \msun.  A handful of the objects identified by these
authors would be visible if present in M 83, and our results are consistent
with this.  \citet{Pisanoetal2007} survey six analogs of the Local
Group and detect no extragalactic \hi\ emission.  Using a similar analysis
to ours, they conclude the Galactic HVCs must be distributed within 90 kpc
and have average \hi\ mass less than about $4\eex{5} \msun$.  The handful
of discrete HVCs we detect in M 83 could represent the high-mass tail of a
much more abundant low-mass, nearby population.

%%%%%%%%%%%%%%%%%%%%%%%%%%%%%%%%%%%%%%%%%%%%%%%%%%%%%%%%%%%%%%%%%%%%%%%%
% SECTION -- Conclusions
%%%%%%%%%%%%%%%%%%%%%%%%%%%%%%%%%%%%%%%%%%%%%%%%%%%%%%%%%%%%%%%%%%%%%%%%
\section{SUMMARY AND CONCLUSIONS}

We have performed deep, wide-field imaging of the \hi\ in M 83, mapping the
outer features for the first time.  The \hi\ ring and warp seen by
\citet{TilanusAllenM83} are confirmed, and the outer arms extend to a
radius of 34 kpc and show reflective symmetry, hinting at a tidal formation
scenario.  At least 80\% of the \hi\ mass is located outside the optical
extent of the galaxy, with the possibility that more \hi\ lies undetected
outside the primary beam HPBW.  The optical companion KK208
\citep[][see Figure \ref{fig:m83opt}]{KarachentsevaKarachentsev98}
contains no \hi\ to a limiting column density of 1\eex{19} \cm.

We discovered a spatially-extended component of anomalous-velocity gas
deviating 40--50 \kps\ from the bulk \hi\ disk and coincident with it in
projection, with a line-of-sight velocity dispersion of 10--15 \kps.  We
interpreted this as a vertically extended disk rotating in the same sense
but about 100 \kps\ more slowly than the kinematically cold, thin disk.
The 5.6\eex{7} \msun\ of \hi\ it contains is 5.5\% of the total \hi\
within the stellar disk of the galaxy.

We have introduced a new technique of searching for faint extended radio
emission, combining several existing tools and including statistically
robust data modeling.  Other recent studies have used similar techniques of
contouring and extended source extraction
\citep{Putmanetal2002,deHeijBraunBurton02}.  These developments are driven
by the need for lower detection limits in projects constraining the missing
satellites problem and obtaining a full census of the Galactic HVCs.

Using this technique, our observations are sensitive to unresolved sources
($d \lesssim 1$ kpc) with masses greater than 5\eex{5} \msun.  We have
discovered 14 discrete anomalous-velocity emission sources, 
of which we consider 8 to be real detections,
ranging in \hi\
mass from 6\eex{5} to 1.5\eex{7} \msun\ and projected on and off the disk.
Three of these sources are high-significance detections and appear to be
distinct \hi\ clouds with masses in excess of 5\eex{6} \msun.  They are
spatially extended and coincide in projection with the optical spiral arms.
The flux distribution of the low-mass sources is inconsistent with purely
spurious detections, therefore we conclude that most of these are real HVCs
in the M 83 system.  They are generally unresolved, off-disk features and
are free of diffuse optical light to a limiting surface brightness of 27
\rsloan\ mag per square arcsec.  

We conclude that a combination of a galactic fountain and tidal stripping
are responsible for the anomalous \hi\ features that we observe in M 83.
The handful of HVCs we have detected are likely to represent the high-mass
end of a substantial HVC population in this galaxy.  We expect future deep
\hi\ observations of nearby spirals to add to the growing number of HVC
analogs detected in external galaxies, and to thereby shed light on the
ubiquity of HVC activity and the nature of our own Galaxy's gaseous
neighborhood.

\acknowledgments

We thank the staff of the VLA and CTIO for their assistance with the
planning and execution of these observations.  We especially thank Jonathan
Williams and Juan Uson for providing access to and help with their source
detection software code, and the anonmyous referee for constructive
comments that improved the manuscript.  EDM would like to thank Mario
Mateo, Hugh Aller, and Tim McKay for their helpful suggestions toward the
improvement of this work.

\appendix

\section{A NEW SOURCE DETECTION METHOD FOR SPECTRAL SYNTHESIS IMAGING}
\label{app:snrch}

We have developed a suite of software that searches a spatially-coherent,
3-d data cube for signal in a statistical way.  This software borrows
heavily from the methods and code developed by \citet{Uson91} and
\citet{WilliamsdeGeusBlitz94}, but as we have made variations to the
methods and translated much of the code to a new programming language, we
here describe the new software in some detail.

% program is SNRCH, written largely in PDL with calls to IDL for the
% CLUMPFIND stuff

The program begins by smoothing the input data cube in velocity with
Gaussian kernels of the form

\begin{equation}
H(v) = e^{-(v^2/2\sigma_{v}^2)},
\label{eq:snrchkernel}
\end{equation}

\noindent
where $\sigma_{v}$ is the width of the particular Gaussian.  The
convolution is performed by a fast Fourier transform.  If one assumes a
Gaussian velocity profile for a clump of emission, the signal due to this
feature as a function of channel or frequency $v$ would be

\begin{equation}
S(v) = A(v_0) e^{-(v-v_0)^2/2\sigma_{v,0}^2},
\label{eq:snrchflux}
\end{equation}

\noindent
and the smoothing function acts as a matched filter, enhancing features of
similar width and de-emphasizing narrower or wider features.  In a given
channel, the normalized amplitude of the smoothed feature is given by

\begin{equation}
\langle A(v_0) \rangle = \frac{\D \int S(v)H(v)dv }
                                {\D \int H(v)^2 dv},
\label{eq:snrchamp}
\end{equation}

\noindent
where the integral is taken over the entire feature.   This value is
maximized when $\sigma_{v} = \sigma_{v,0}$.   The normalization factor
$1/{\D \int H(v)^2 dv}$ is such that the amplitude of the smoothed feature
will equal the amplitude of the original feature under this same condition;
otherwise, the smoothed amplitude will be depressed compared to the
unsmoothed value.

One can calculate the significance level of such a detection by recognizing
that the smoothed noise is 

\begin{equation}
\langle \sigma(v_0)^2 \rangle = 
     \frac{\D \int \sigma_i(v)^2 H(v-v_0)^2 dv }
          {\D \left[ \int H(v)^2 dv \right] ^2}\;\;\;,
\label{eq:snrchnoise}
\end{equation}

\noindent
if we assume that neighboring channels are independent and the noise (given
as $\sigma_i$ for channel $i$) sums in quadrature.  The noise in each
channel is determined by reflecting the negative portion of the data
histogram across the mean and performing an iterative 4-$\sigma$ rejection
until the measured $\sigma$ converges.  
The significance level (S/N) is calculated as $\langle A(v_0) \rangle /
\langle \sigma(v_0)^2 \rangle^{1/2}$ and written to disk as a data cube.

Peaks in S/N are identified down to some detection threshold, typically
around 4-$\sigma$, with neighboring above-threshold pixels merged into
groups.  These 3-d groups or ``islands'' serve as kernels for an iterative
contouring scheme, with successively lower thresholds applied to the S/N
cube.  Each group is extended by adding new pixels that are above threshold
and share at least two corners with any member pixel.  That is, only
neighbors that share at least one dimensional plane with the member are
added, so that neighbors at a diagonal in all three dimensions are
excluded.  The algorithm for this procedure is based on the CLUMPFIND
software developed by \citet{WilliamsdeGeusBlitz94}.  The contours are
produced in intervals (which only affect the processing speed) down to a
specified level, typically 3-$\sigma$, iterating at each level until no new
neighbors are added to any groups and merging groups which meet the
neighbor criterion.  Finally, any immediate neighbors (i.e., sharing four
corners or two parameter planes) above 1.5-$\sigma$ are added to each group
to improve flux measurements.  No merging or iterating is done at this
point.  Groups are eliminated if they contain fewer pixels than a
completely unresolved source (10 pixels in our case), as these features are
unphysical. 

The group IDs are written to a data cube, which is applied to the original
data cube as a mask for each group in turn.  The total flux, centroid, and
other parameters of the group are determined from the original data cube,
using only the pixels that are members of the group.  The primary beam
correction is applied at this point, and the corrected flux and mass are
recorded along with the uncorrected parameters.

\end{document}